\documentclass[manuscript]{aastex}

\usepackage{graphicx}
\usepackage{float}
\usepackage{amsmath}
\usepackage{natbib}
\usepackage{threeparttable}

\shorttitle{The Mass of Kepler-21b}
\shortauthors{L$\'o$pez-Morales et al.}

\begin{document}



\title{Kepler-21$\MakeLowercase{{\rm b}}$: A rocky planet around a ${\rm V =}$ 8.25 magnitude star\footnotemark[1]}


\author{Mercedes L${\rm \acute{o}}$pez-Morales\altaffilmark{2}, Rapha${\rm \ddot{e}}$lle D. Haywood\altaffilmark{2}, Jeffrey L. Coughlin\altaffilmark{3}, Li Zeng\altaffilmark{4}, Lars A. Buchhave\altaffilmark{5}, Helen A. C. Giles\altaffilmark{6}, Laura Affer\altaffilmark{7},  Aldo S. Bonomo\altaffilmark{8}, David Charbonneau\altaffilmark{2}, Andrew Collier Cameron\altaffilmark{9}, Rosario Cosentino\altaffilmark{10}, Courtney D. Dressing\altaffilmark{11}, Xavier Dumusque\altaffilmark{6}, Pedro Figueira\altaffilmark{12}, Aldo F. M. Fiorenzano\altaffilmark{10}, Avet Harutyunyan\altaffilmark{10}, John Asher Johnson\altaffilmark{2}, David W. Latham\altaffilmark{2}, Eric D. Lopez\altaffilmark{13}, Christophe Lovis\altaffilmark{6}, Luca Malavolta\altaffilmark{14,15},  Michel Mayor\altaffilmark{6}, Giusi Micela\altaffilmark{7}, Emilio Molinari\altaffilmark{10,16}, Annelies Mortier\altaffilmark{9}, Fatemeh Motalebi\altaffilmark{6}, Valerio Nascimbeni\altaffilmark{14}, Francesco Pepe\altaffilmark{6}, David F. Phillips\altaffilmark{2}, Giampaolo Piotto\altaffilmark{14,15}, Don Pollacco\altaffilmark{17}, Didier Queloz\altaffilmark{6,18}, Ken Rice\altaffilmark{13}, Dimitar Sasselov\altaffilmark{2}, Damien Segransan\altaffilmark{6},  Alessandro Sozzetti\altaffilmark{8}, Stephane Udry\altaffilmark{6}, Andrew Vanderburg\altaffilmark{2}, Chris Watson\altaffilmark{19}}

\email{mlopez-morales@cfa.harvard.edu}

\altaffiltext{2}{Harvard-Smithsonian Center for Astrophysics, 60 Garden Street, Cambridge, MA 01238, USA}
\altaffiltext{3}{SETI Institute, 189 Bernardo Avenue Suite 200, Mountain View, CA 94043, USA}
\altaffiltext{4}{Department of Earth and Planetary Sciences, Harvard University, 20 Oxford Street, Cambridge, MA 01238, USA}
\altaffiltext{5}{Centre for Star and Planet Formation, Natural History Museum of Denmark \& Niels Bohr Institute, University of Copenhagen, DK-1350 Copenhagen, Denmark}
\altaffiltext{6}{Observatoire Astronomique de l'Universit${\rm \acute{e}}$ de Gen${\rm \acute{e}}$ve, Chemin des Maillettes 51, Sauverny, CH-1290, Switzerland}
\altaffiltext{7}{INAF - Osservatorio Astronomico di Palermo, Piazza del Parlamento 1, 90134 Palermo, Italy}
\altaffiltext{8}{INAF - Osservatorio Astrofisico di Torino, via Osservatorio 20, 10025 Pino Torinese, Italy}
\altaffiltext{9}{SUPA, School of Physics $\&$ Astronomy, University of St.~Andrews, North Haugh, St. Andrews Fife, KY16 9SS, UK}
\altaffiltext{10}{INAF - Fundaci${\rm \acute{o}}$n Galileo Galilei, Rambla Jos${\rm \acute{e}}$ Ana Fernandez P${\rm \acute{e}}$rez 7, 38712 Bre${\rm \tilde{n}}$a Alta, Spain}
\altaffiltext{11}{NASA Sagan Fellow, Division of Geological and Planetary Sciences, California Institute of Technology, Pasadena, CA
91125, USA}
\altaffiltext{12}{Instituto de Astrofisica e Ciencias do Espaco, Universidade do Porto, CAUP, Rua das Estrelas, PT4150-762 Porto, Portugal}
\altaffiltext{13}{SUPA, Institute for Astronomy, University of Edinburgh, Royal Observatory, Blackford Hill, Edinburgh, EH93HJ, UK}
\altaffiltext{14}{Dipartimento di Fisica e Astronomia "Galileo Galilei", Universita'di Padova, Vicolo dell'Osservatorio 3, 35122 Padova, Italy}
\altaffiltext{15}{INAF - Osservatorio Astronomico di Padova, Vicolo dell'Osservatorio 5, 35122 Padova, Italy}
\altaffiltext{16}{INAF - IASF Milano, via Bassini 15, 20133, Milano, Italy}
\altaffiltext{17}{Department of Physics, University of Warwick, Gibbet Hill Road, Coventry CV4 7AL, UK}
\altaffiltext{18}{Cavendish Laboratory, J J Thomson Avenue, Cambridge CB3 0HE, UK}
\altaffiltext{19}{Astrophysics Research Centre, School of Mathematics and Physics, Queens University, Belfast, Belfast BT7 1NN, UK}

\footnotetext[1]{Based on observations made with the Italian Telescope Nazionale Galileo (TNG) operated on the island of La Palma by the Fundacion Galileo Galilei of the INAF (Istituto Nazionale di Astrofisica) at the Spanish Observatorio del Roque de los Muchachos of the Instituto de Astrofisica de Canarias.}

\begin{abstract}
HD 179070, ${\it aka}$ Kepler-21, is a ${\rm V}$ = 8.25 F6IV star and the brightest exoplanet host  discovered by ${\it Kepler}$. An early detailed analysis by \cite{Howell2012} of the first thirteen months (Q0 -- Q5) of ${\it Kepler}$ light curves revealed transits of a planetary companion, Kepler-21b, with a radius of about 1.60 $\pm$ 0.04 ${\rm R_{\oplus}}$ and an orbital period of about 2.7857 days. However, they could not determine the mass of the planet from the initial radial velocity observations with Keck-HIRES, and were only able to impose a 2$\sigma$ upper limit of 10 ${\rm M_{\earth}}$. Here we present results from the analysis of 82 new radial velocity observations of this system obtained with HARPS-N, together with the existing 14 HIRES data points.  We detect the Doppler signal of Kepler-21b with a radial velocity semi-amplitude ${\rm K}$ =  2.00 $\pm$ 0.65 ${\rm m~s^{-1}}$, which corresponds to a planetary mass of 5.1 $\pm$ 1.7 $\rm M_{\oplus}$. We also measure an improved radius for the planet of 1.639$^{\rm +0.019}_{\rm -0.015}$ $\rm R_{\oplus}$, in agreement with the radius reported by \cite{Howell2012}. We conclude that Kepler-21b, with a density of 6.4 $\pm$ 2.1 ${\rm g~cm^{-3}}$,  belongs to the population of terrestrial planets with iron, magnesium silicate interiors, which have lost the majority of their envelope volatiles via stellar winds or gravitational escape. The radial velocity analysis presented in this paper serves as example of the type of analysis that will be necessary to confirm the masses of TESS small planet candidates.

\end{abstract}





\section{Introduction}

Results from NASA's ${\it Kepler}$ Mission have revealed an abundance of planets smaller than 2 ${\rm R_{\earth}}$ with orbital periods less than 100 days \citep{Howard2012,Fressin2013,Dressing2013,Petigura2013a,Petigura2013b,ForemanMackey2014,Silburt2015,Dressing2015}. Although only a few of those planets have measured masses, and therefore densities, those measurements have started to unveil an interesting picture. Below a radius of about 1.6 ${\rm R_{\earth}}$ most planets are consistent with bare rocky compositions without any significant volatile envelopes \citep{Rogers2015}. Moreover, when considering only planets with masses measured with precisions better than 20$\%$ via radial velocities, planets with masses smaller than about 6 ${\rm M_{\earth}}$ appear to be rocky and have interiors composed mostly of iron and magnesium silicates in Earth-like abundances \citep[26$\%$ Fe, 74$\%$ ${\rm MgSiO_3}$, on average, based on][]{Zeng2016}, while planets more massive than about 7 ${\rm M_{\earth}}$  show a wider range of densities \citep{Dressing2015b,Gettel2016,Buchhave2016}. Such a dichotomy suggests the possible existence of mechanisms by which planets more massive than approximately 7 ${\rm M_{\earth}}$ in orbits of only a few days can retain significant volatile envelopes, while less massive planets lose all the material in their outer layers to a combination of the effect of stellar winds and atmospheric escape. 

However, despite the rapid observational progress on the determination of fundamental properties of low mass planets, some basic questions about the origin of this short-period rocky planet population are still not understood. Almost all of the confirmed rocky planets are on highly irradiated orbits, where they are bombarded by large amounts of ionizing EUV and X-ray radiation, which can drive a photo-evaporative wind from the atmosphere of the planet and over a planet's lifetime can remove a significant amount of mass from planets with volatiles envelopes \citep[e.~g.][]{Owen2012}. Several recent studies have shown that Kepler's short-period super-Earths and sub-Neptunes have likely been significantly sculpted by photo-evaporation \citep[e.~g.][]{Lopez2012,Lopez2013,Owen2013}, or else by some other comparable process like atmospheric erosion by impacts \citep[e.~g.][]{Inamdar2015,Schlichting2015}. Thus, while it is possible that the short-period rocky planets simply formed with their current Earth-like compositions, their low masses and highly irradiated orbits mean that they could also be the remnant cores of volatile-rich hot Neptunes which have lost their envelopes. Even considering all these scenarios, it is not clear why a transition between bare cores and planets with significant volatiles would occur at 1.6 $\rm R_{\earth}$. For example, recent precise mass measurements of planets with masses between 3 and 8 ${\rm M_{\earth}}$ and periods out to 17 days, via transit timing variations, reveal a wide range of densities for planets with masses near 5--6 ${\rm M_{\earth}}$, analogous to the situation for more massive planets \citep{JontofHutter2016}. The recently discovered Kepler-20b, with a mass of 9.7 ${\rm M_{\earth}}$, radius 1.9 $\rm R_{\earth}$, and a orbital period of 3.7 days appears to be a bare core \citep{Buchhave2016}.

With the current sample of small planets with precise mass measurements it is not  possible to establish whether stellar irradiation is the cause of the bare core to volatile rich envelopes transition. It is also not possible to discern whether the transition is abrupt or smooth \citep{Rogers2015}. We therefore need a larger number of precise mass measurements, especially around the apparent 1.6 $\rm R_{\earth}$ transition region.

In this paper we report a new mass determination for Kepler-21b, a 5.1 $\pm$ 1.7 ${\rm M_{\earth}}$ super-Earth 
located near the apparent  mass boundary between predominately volatile-poor super Earths and volatile-rich larger planets. Kepler-21b orbits the brightest exoplanet host star discovered by ${\it Kepler}$ (HD 179070, V = 8.25), which is also a slightly evolved F6IV star. An earlier  study of this planet by \cite{Howell2012}, based on the first six quarters of ${\it Kepler}$ data (Q0--Q5),  found a planet radius of 1.6 $\pm$ 0.04 $\rm R_{\earth}$, but could not determine the planetary mass because of the effect of the stellar variability on the radial velocity (RV) measurements. Our mass measurement comes from new radial velocity data collected with HARPS-N between 2014 and 2015, combined with the HIRES data from  \cite{Howell2012} and fitted using Gaussian Processes regressions (GPs). In addition, we compute a new planetary radius from the complete ${\it Kepler}$ Q0-Q17 light curves, detrended from stellar variability using new time series analysis techniques.  

We describe the light curve and radial velocity analyses in Section 2. In Sections 3 and 4, we describe the light curve and radial velocity fits and their results. Finally, we discuss our findings and summarize our conclusions in Section 5.

\section{Data}

\subsection{Kepler Photometry} \label{sec:kepler}

\begin{figure}[t]
\centering
\includegraphics[scale=0.6]{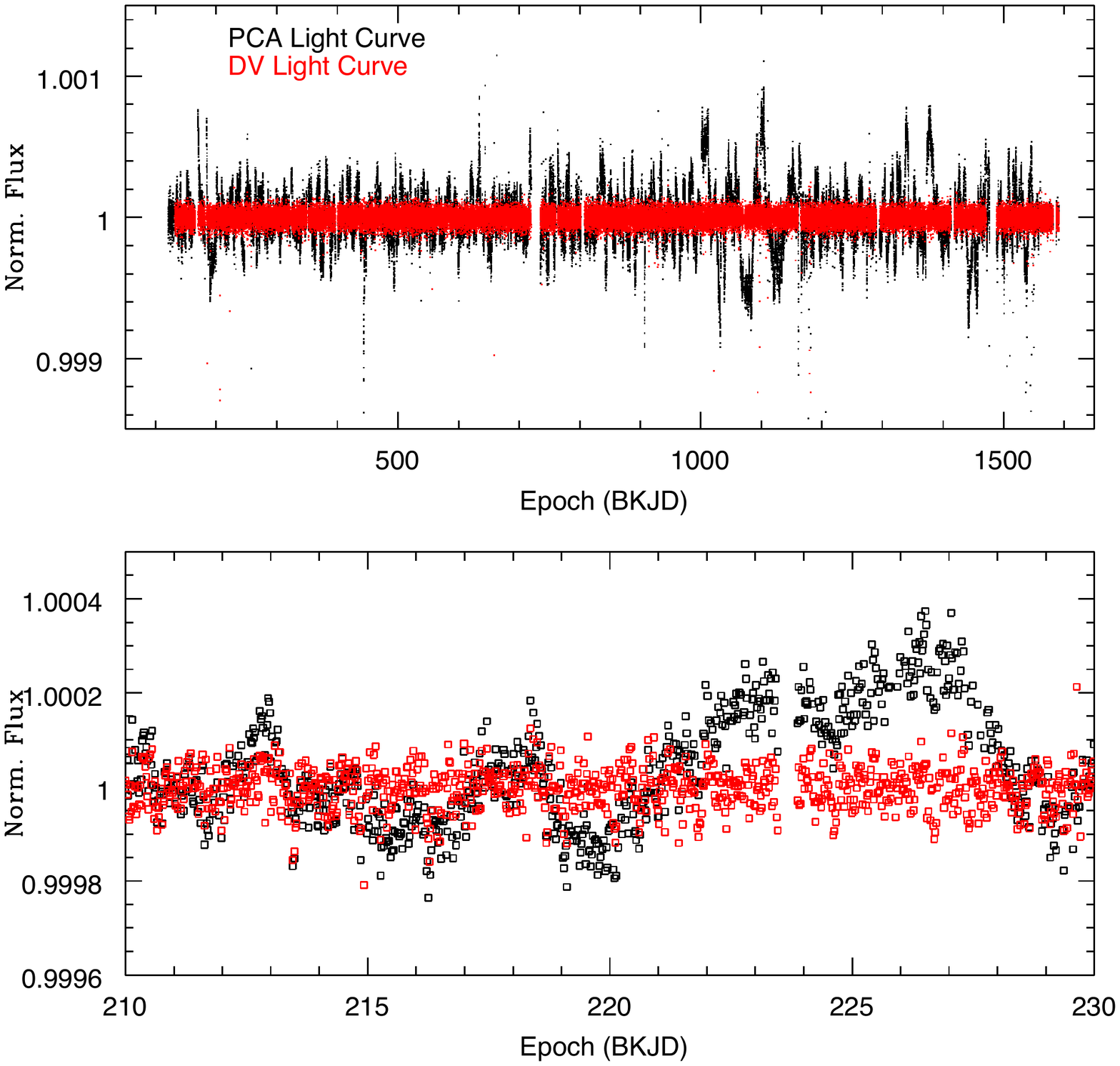}
\caption{{\it Top} -- Full Q0--Q17 Kepler light curve analyzed using Data Validation (DV; red) and Principal Component Analysis (PCA; black) detrendings.  {\it Bottom} -- A 20-day zoom-in of the light curves to illustrate the difference between the DV and the PCA detrendings, this last one preserving the variability signal of the star. The times are given in Kepler Barycentric Julian Dates (BKJD), i.e. BJD - 2454833.0.\label{fig:sampleLCs}}
\end{figure}

Kepler-21 was monitored with ${\it Kepler}$ in 29.4 min, long cadence mode between quarters Q0 and Q17, and in 58.9 sec, short cadence mode in quarters Q2 and Q5-Q17, covering a total time period of 1,470.5 days (BJD 2454953.540 -- 2456424.002). We analyzed the full ${\it Kepler}$ dataset using two different detrendings: Data Validation (DV) and Principal Component Analysis (PCA). The results of both analyses are shown in Figure~\ref{fig:sampleLCs}.

The result of the DV analysis is the detrended flux time-series available in the DV report summaries \citep{Wu2010} as obtained from the NASA Exoplanet Archive's Q1-Q17 DR24 TCE table\footnote{http://exoplanetarchive.ipac.caltech.edu}. For this detrending, as detailed in \cite{Jenkins2010}, an optimal photometric aperture is used to sum up the flux from the central pixels of the image and produce a time-series light curve. The Pre-Search Data Conditioning (PDC) module then removes systematic trends that are common to multiple stars on the detector. The resulting time-series is then run through a harmonic filter that identifies and removes sinusoidal trends in the data. Finally, a median detrender is used to remove any remaining photometric variations at durations longer than the transit duration and normalize the data \citep[see][for more details]{Wu2010}.

While the DV detrending produces a very clean light curve, any variations at timescales greater than the transit duration, such as the planet's phase curve or stellar variations due to rotation or pulsation, are removed. This is due mainly to the harmonic filter and median detrender, which is selected to preserve features with timescales of the order of the transits. Detailed inspection of the light curves also shows that the PDC module significantly suppresses sinusoidal-like astrophysical signals at 10 days, and completely removes them by 20 days \citep{Christiansen2013}. Stellar rotation periods, which can be confused with an exoplanet's radial velocity signal, are usually in that same 10--20 day period range \citep[see e.~g.][]{McQuillan2014}, so it is important to preserve the stellar signal. Therefore, we also employed a PCA detrending \citep{Murtagh1987}, similar to that described in \cite{Coughlin2012}\footnote{This tool is now publicly available as a task called ${\it keppca}$ in the Kepler PyKE tools package (http://keplergo.arc.nasa.gov/PyKE.shtml)}. For the PCA detrending, all available pixels in the image are summed up to produce a time-series light curve. A PCA is then run on the pixel-level time-series data to obtain a series of basis vector components. These components correspond to the systematic trends belonging to the specific target being analyzed that arise due to motion on the detector, as well as instrumental variation and cosmic ray impacts. These basis vectors are removed from the time-series photometry, which is then normalized by simply dividing by the median flux level in each quarter. The advantage of this PCA detrending is that it preserves the intrinsic photometric signals introduced by both the star and the planet, while removing systematic trends from the spacecraft and detector.

Although Kepler-21, at ${\rm V}$ = 8.25, is saturated on the detector, both detrendings include all the saturated pixels. Since charge is conserved on the ${\it Kepler}$ CCDs to a very high degree, accurate differential photometry is achievable for saturated objects, as long as enough pixels are included to capture all the saturated regions and a significant amount of the star's point spread function \citep{Koch2010}.

\subsection{HARPS-N Spectroscopy}\label{sec:harpsn}

We collected a total of 82 radial velocity (RV) observations of Kepler-21 with the HARPS-N spectrograph installed on the 3.6-m Telescopio Nazionale Galileo (TNG) at the Observatorio del Roque de los Muchachos in La Palma, Spain \citep{Cosentino2012}. HARPS-N is an updated version of HARPS at the ESO 3.6-m \citep{Mayor2003}, and has already produced a series of high-precision RV results \citep[e.g. ][Buchhave et al. 2016]{Covino2013,Pepe2013,Bonomo2014,Desidera2014,Dumusque2014, Esposito2014,LopezMorales2014,Damasso2015,Dressing2015b,Mancini2015,Motalebi2015,Sozzetti2015,Gettel2016,Malavolta2016}. 

We observed Kepler-21 between April 2014 and June 2015 as part of the HARPS-N Collaboration's Guaranteed Time Observations (GTO) program, following a standard observing approach of one or two observations per night, separated by 2--3 hours, on nights assigned to the GTO program. Kepler-21 is a bright target with ${\rm V}$ = 8.25 ($K_{\rm p}$ = 8.2), so we obtained spectra with signal-to-noise ratios in the range SNR = 45 -- 308 (average SNR = 167), at 550 nm in 10 -- 30 minute exposures, depending on the seeing and sky transparency.  A summary of the observations is provided in Table~\ref{tab:data}.

The average RV error of the observations is 1.59 $\pm$ 0.68 ${\rm m~s^{-1}}$. This value is larger than the expected error of about 1.00 ${\rm m~s^{-1}}$ for a slowly rotating F- or G-dwarf of similar apparent magnitude, but we attribute it to the faster rotation of this star ($v\,\sin\,i_\star$ = 8.4  km s$^{-1}$; see section~\ref{sec:vsini}), which broadens the spectral lines and therefore gives a larger uncertainty on the RV determination. In addition, Kepler-21 presents significant photometric and spectroscopy variability, which produces an observed radial velocity variation semiamplitude of about 10 ${\rm m~s^{-1}}$, including the stellar and planetary signals.

\begin{figure}
\centering
\includegraphics[scale=0.5]{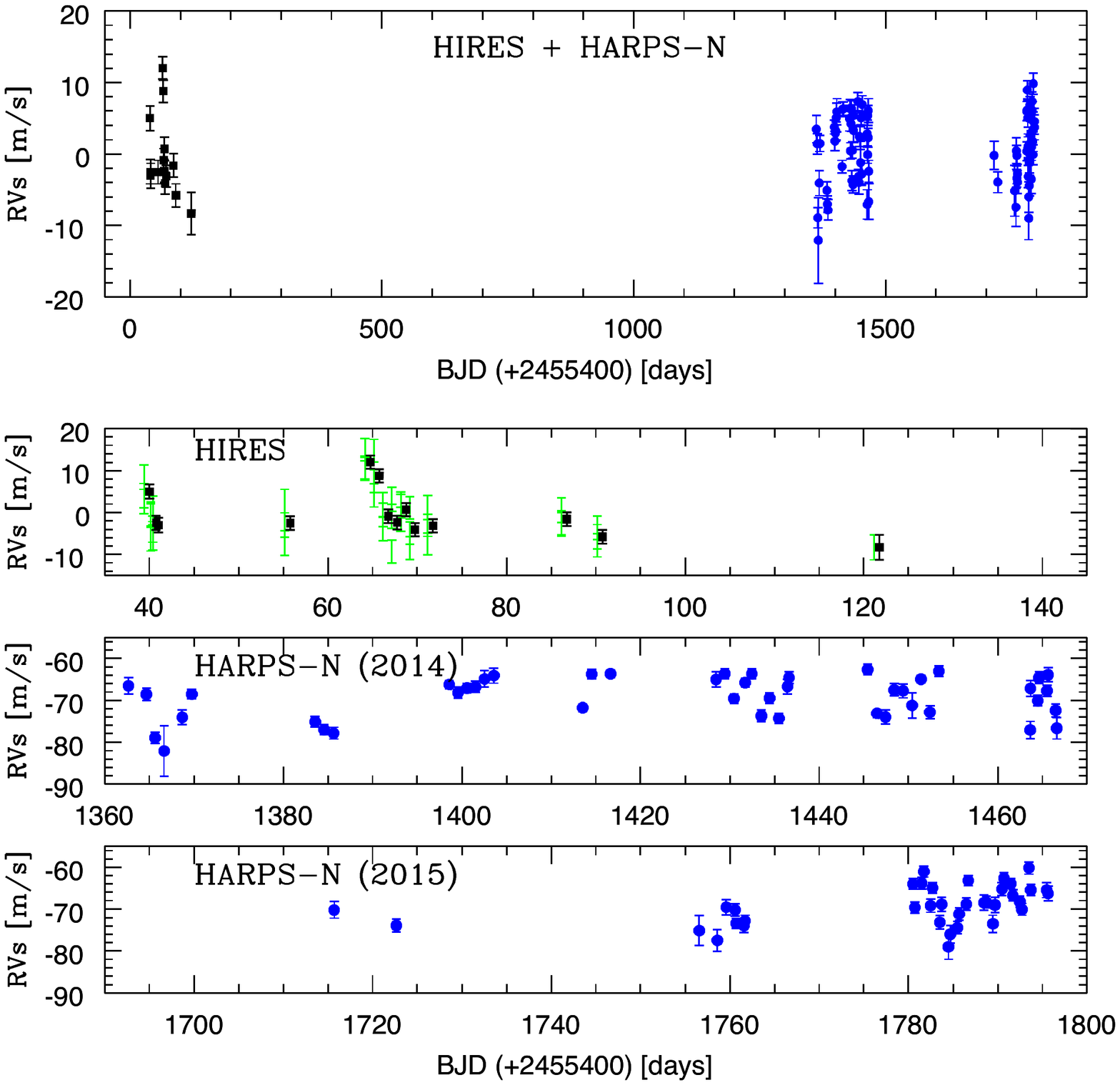}
\caption{The top diagram shows the combined HIRES (black squares) and HARPS-N (blue circles) radial velocity observations corrected from a systemic velocity for the star of 19100 $\rm m~s^{-1}$ and for offsets between the datasets. The bottom three diagrams show the HIRES data and the HARPS-N 2014 and 2015 data separately, plotted over the same timescale (110 days). The green lines in the HIRES plot show the individual 2--3 min observations. The black squares show the weighted average of each set of three consecutive observations, as reported by \cite{Howell2012}. The green lines have been shifted by  -0.6 BJDs for clarity. The difference in the vertical scales between the HIRES and the HARPS-N plots shows the instrumental offset of 70 m s$^{-1}$ between both datasets.\label{fig:allrvs}}
\end{figure}

The spectra were reduced with version 3.7 of the HARPS-N Data Reduction Software (DRS), which includes corrections for color systematics introduced by variations in seeing \citep{Cosentino2014}. The radial velocities were computed using a numerical weighted mask following the methodology outlined by \citet{Baranne1996}. The resultant radial velocities are presented in Table~\ref{tab:data} and in Figure~\ref{fig:allrvs}. Table~\ref{tab:data} also includes each observation's central BJD, exposure time, bisector span and the measured log $\rm R'_{HK}$ activity index.

\subsection{HIRES Spectroscopy}\label{sec:hires}

\cite{Howell2012} published 14 radial velocity observations of Kepler-21 collected between August 31 and November 21 2010 with HIRES on Keck. They adopted a different observing strategy than the one we used with HARPS-N: they observed the target in groups of three consecutive exposures, each lasting between 2 and 3 min in order to maintain a typical SNR of about 210 and an internal RV error of about 2 ${\rm m~s^{-1}}$ per exposure. The sampling of each group of exposures varied from twice a night to once every fifteen days. In total they collected 13 groups of three exposures in this manner over a period of 51 days and a final single exposure 31 days later.  We show the HIRES RVs together with the new HARPS-N RVs in Figure~\ref{fig:allrvs}.

\section{Analysis of the Photometric and Radial Velocity Data}

\subsection{Preliminary Analysis of the LCs}

\begin{figure}[t]
\centering
\includegraphics[scale=0.45]{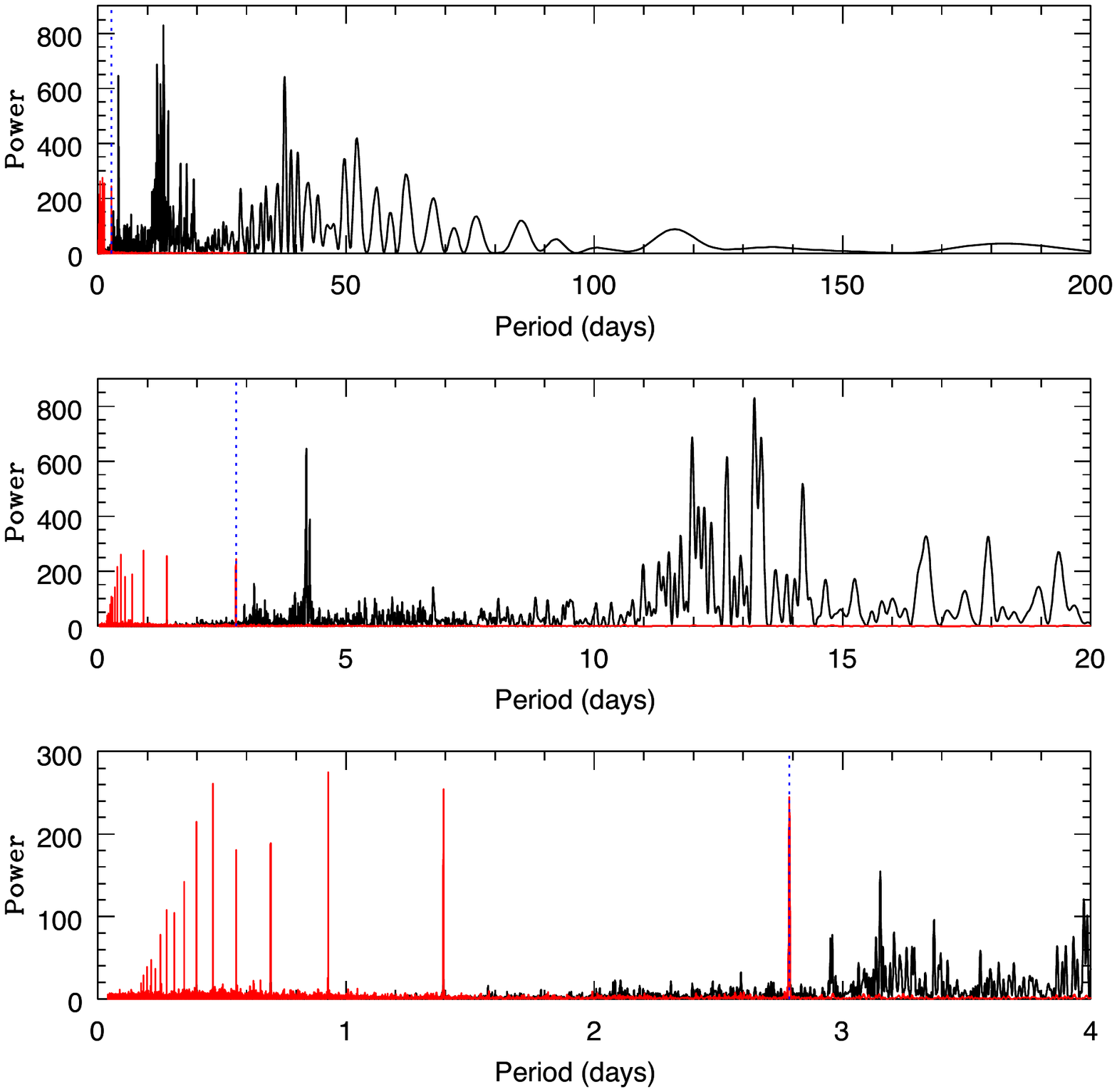}
\caption{Generalized Lomb-Scargle periodograms of the DV light curve (red) and the PCA light curve (black). The vertical dashed, blue line indicates the period of Kepler-21b (P = 2.7858d).\label{fig:LCPeriodogram}}
\end{figure}

From the DV and the PCA light curve analyses shown in Figure~\ref{fig:sampleLCs}, the PCA light curve, which preserves the variability signal from the star, reveals stellar variability with a standard deviation of 145 ppm and peak to peak variations of about 1300 ppm. In the DV light curve, after eliminating in-transit points, the standard deviation of the light curve baseline gets reduced to 53 ppm. Figure~\ref{fig:LCPeriodogram} shows superimposed Generalized Lomb-Scargle \citep[GLS,][]{Zechmeister2009} periodograms of the PCA and DV light curves. In the case of the PCA light curve several strong peaks with P $\lesssim$ 50 days dominate the periodogram. The strongest peak is at a period of 13.25 days, with several other strong peaks near that value. There are also strong, isolated peaks at 4.2 and 37.7 days, which are 1/3 and 3 times the rotation period of the star found using autocorrelation functions, as detailed below. The orbital period of Kepler-21b is not visible in the PCA light curve's periodogram, however, the periodogram of the DV light curve, where stellar variability has been removed and only the planetary transits remain, shows clearly a peak at a period of 2.7858 days, and its harmonics (e.g. 1.39 and 0.92 days). There is no other significant peak in the DV light curve periodogram.

\begin{figure}[b]
\centering
\includegraphics[scale=0.45]{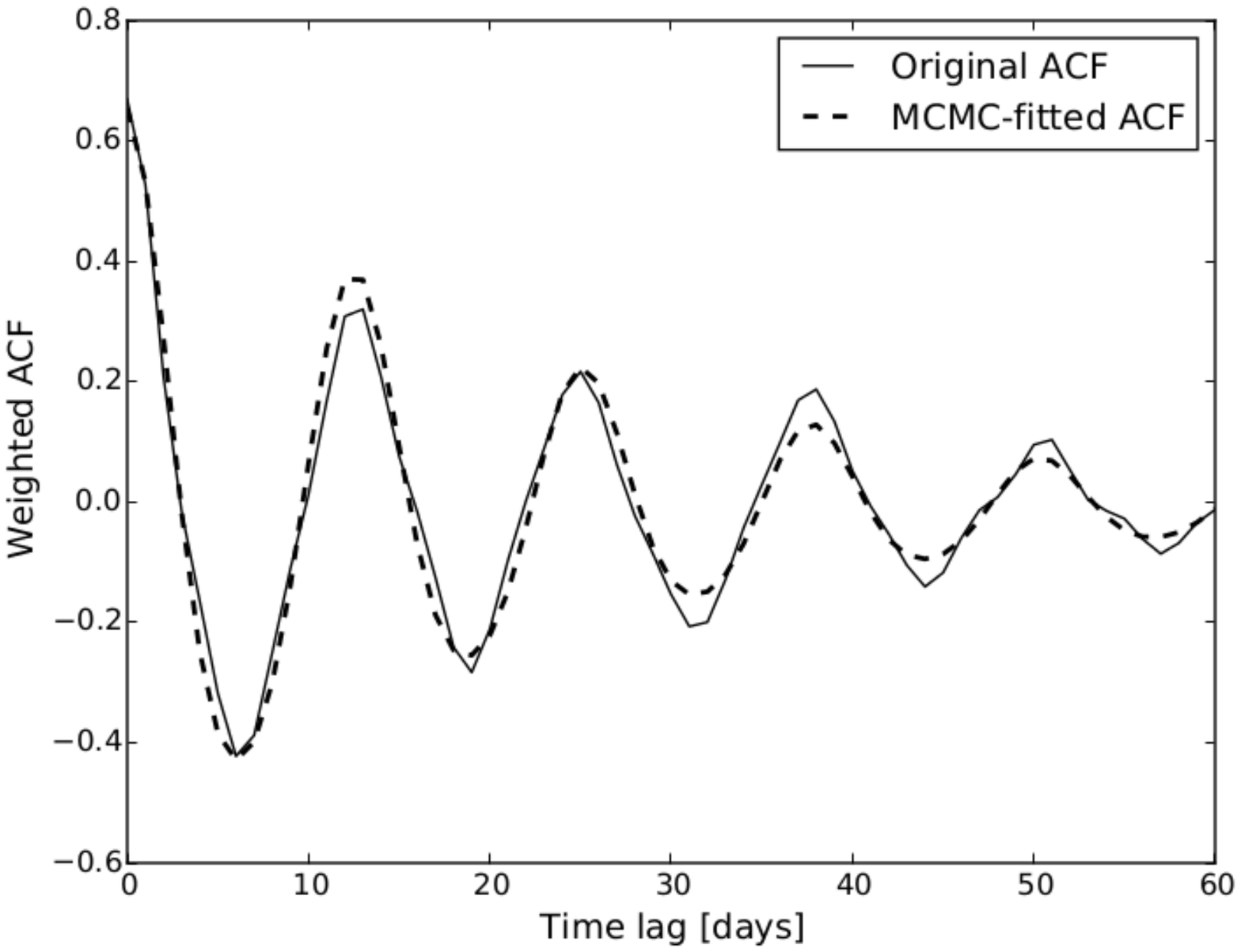}
\caption{Autocorrelation function of the Kepler-21 light curve (solid line), and the resulting MCMC model of that function using an underdamped simple harmonic oscillator (dashed line).\label{fig:ACFplot}}
\end{figure}

To obtain a better estimation of the stellar rotation period and measure the star spot decay times we applied an autocorrelation function (ACF) to the PCA light curve.
We produced the ACF by introducing discrete time lags, as described by \cite{Edelson1988}, in the light curve and cross-correlating the shifted light curves with the original, unshifted curve. The result is illustrated in Figure~\ref{fig:ACFplot}.
The ACF peaks in the figure correspond to time offsets that coincide with an integer multiple of the rotation period of the star. In addition, the effective decay time of the spots can be estimated by measuring the amplitude decay of the ACF side lobes in the figure. The amplitude decay occurs as the spots fade away with time. To measure these two parameters, we fitted the positive ACF lobes to the equation of motion for an underdamped simple harmonic oscillator (uSHO), which has a similar shape to the ACF shape. However, it has been also found that a large number of stars exhibit {\it interpulses}, which occur when an additional large spot appears on the opposite side of the star, introducing additional side lobes at half periods (Giles \& Cameron, in prep). This can be accounted for by introducing an additional cosine term in the uSHO equation. Therefore, the uHSO equation used here has the form 
\begin{equation}
y = e^{-At} (B \cos{\omega t} + C \cos{2 \omega t}) + y_0 , 
\end{equation}

\noindent where A is the spot decay timescale of the ACF, in ${\rm days^{-1}}$, and $\rm \omega$ is the frequency, also in ${\rm days^{-1}}$. B and C are coefficients representing the amplitudes of the cosine terms and $\rm y_0$ is an offset term from $\rm y=0$.

We fit the uSHO equation to the ACF using a Monte-Carlo-Markov-Chain (MCMC) method, with starting parameters determined from
the ACF, and step sizes drawn from a Gaussian distribution with parameter errors as the variance. The MCMC was performed
twice: first to find the highest likelihood values; and second to explore that likelihood peak for the optimum set of
values. The errors and the step size in the second MCMC were refined using the variance of the last 5000 steps in the first MCMC fit.
Convergence was reached when the median of all previous likelihood values was greater than the current likelihood
\citep{Charbonneau2008,Knutson2008}. Using this technique, we find a stellar rotation period of 12.62 $\pm$ 0.03 days  and a spot decay time of 24.0 $\pm$ 0.1 ${\rm days^{-1}}$. We notice that this stellar rotational period is slightly shorter than the 13.25 day period found using a GLS periodogram. As shown in Figure~\ref{fig:LCPeriodogram}, there is a set of strong peaks between 11 and 15 days in the GLS periodogram of the PCA curve. That set of peaks is consistent with a period of 12.6 days, and we attribute the difference between the GLS and the ACF results to the spot decay time, which is not accounted for in the GLS periodogram, and the long time baseline of the ${\it Kepler}$ light curve, which likely includes many different, evolving spot configurations emerging at different rotation phases. We adopt the period of 12.6 days found by the ACF analysis as most reliable.

\subsection{Preliminary Analysis of the RV curves}

\begin{figure}[t]
\centering
\includegraphics[scale=0.45]{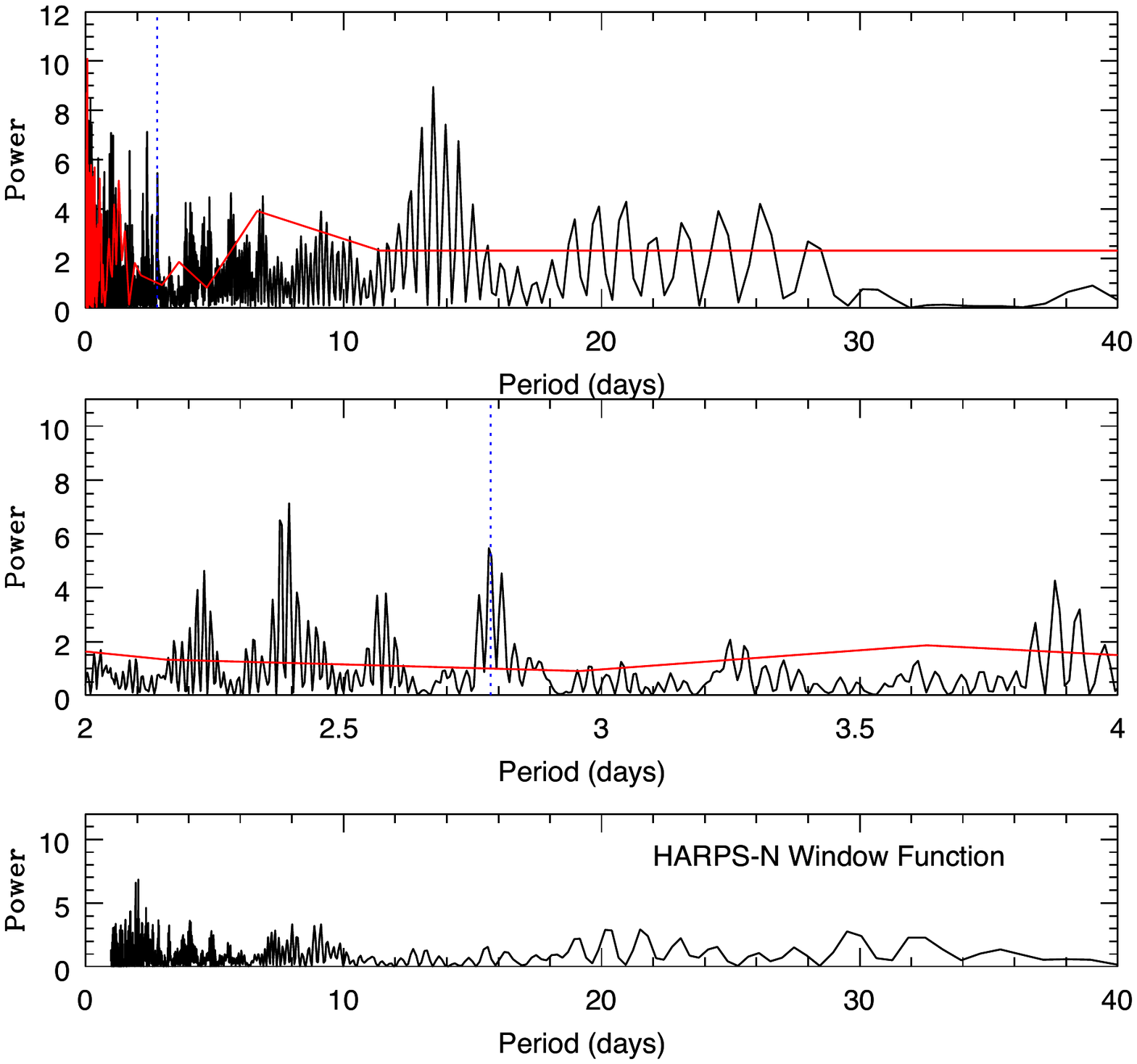}
\caption{{\it Top, Middle} -- Generalized Lomb-Scargle periodograms of the HIRES (red) and the HARPS-N (black) radial velocities. The vertical dashed, blue line indicates the period of Kepler-21b (P = 2.7858d). The signal from the planet is visible in the HARPS-N RVs, but not in the HIRES RVs. {\it Bottom}: -- Window function GLS periodogram of the HARPS-N RVs. \label{fig:RVsPeriodograms}}
\end{figure}

A GLS periodogram analysis of the HARPS-N radial velocities also reveals a complicated structure of peaks, as illustrated in Figure~\ref{fig:RVsPeriodograms}. In this case, the periodogram of the RVs shows a peak at the orbital period of Kepler-21b, but it is not the most significant peak. The strongest peak is at 13.47 days and there is a set of smaller peaks centered around that peak with similar structure to those in the PCA light curve periodogram in Figure~\ref{fig:LCPeriodogram}. We investigated whether that peak structure was produced by the observational window function, with negative result (see bottom panel of Fig.~\ref{fig:LCPeriodogram}). We conclude that this peak is the same as the one observed in the light curve at 13.25 days. Both peaks, and the piramid-shaped structures of other strong peaks around them, result from splitting  of the rotational modulation peak arising from phase and amplitude changes as active regions grow and decay over the long time baseline of the observations.
 
\noindent Combining the 12.6 day period derived from the ACF analysis of the light curve with the 14.83  $\pm$ 2.41 day period derived in section~\ref{sec:vsini} we estimate a inclination for the spin axis for the star $i_\star$ = $58^{+32}_{-11}$ degrees. With that result we cannot confirm a star-planet misalignment. The Generalized Lomb-Scargle periodogram of the HIRES RVs, shown in red in Figure~\ref{fig:RVsPeriodograms}, reveals no significant peak at the period of the planet, consistent with the non-detection reported by \cite{Howell2012}.

\subsection{Stellar Parameters, Rotation and Activity}\label{sec:vsini}

Using the large number of high SNR, high-resolution spectra gathered by HARPS-N we re-determined the stellar parameters of Kepler-21 using the Stellar Parameter Classification pipeline \citep[SPC;][]{Buchhave2014}, and the ARES+MOOG method described in \cite{Mortier2014}. We analyzed 78 of the 82 spectra with exposure times larger than 900 seconds and a resolution of R = 115,000 resulting in an average SNR per resolution element of 300 in the  Mg B line region.  The remaining four spectra not included in the analysis had either lower SNR because of shorter exposure times, or some artifact in the MgB region. The stellar parameter values we obtain with SPC are  ${\rm T_{eff}}$=6216 $\pm$ 50 K and ${\rm [Fe/H]}$= -0.06 $\pm$ 0.08, when using the asteroseismic value of ${\rm log(g)}$ = 4.019 $\pm$ 0.009 derived by \cite{Howell2012} as a prior. Adopting the more recent  asteroseismic value of ${\rm log(g)}$ = 4.026 $\pm$ 0.004 derived by \cite{SilvaAguirre2015} gives similar results. Leaving ${\rm log(g)}$ as a free parameter in our fit yields ${\rm log (g)}$ = 3.87 $\pm$ 0.10, slightly lower than the value reported by \cite{Howell2012} and \cite{SilvaAguirre2015}, and a ${\rm T_{eff}}$=6127 $\pm$ 49 K and ${\rm [Fe/H]}$= -0.11 $\pm$ 0.08.  The parameter values obtained with the ARES+MOOG method are all consistent with the values from SPC. We notice that the ${\rm log(g)}$ derived from the HIRES spectra in Table 3 of \cite{Howell2012} also favors a lower value than the one yielded by the asteroseismology analysis.
However, it has been previously shown that spectroscopic analyses are affected by degeneracies in ${\rm log(g)}$, ${\rm T_{eff}}$, and ${\rm [Fe/H]}$, which generally result in an underestimation of ${\rm log(g)}$ \citep{Torres2012}. In addition, Kepler-21, being a bright star, has a Hipparcos parallax measurement of 8.86 $\pm$ 0.58 ${\rm mas}$ \citep{VanLeeuwen2007}, which corresponds to a distance for the system of 113 $\pm$ 7 ${\rm pc}$ and a stellar radius of 1.96 $\pm$ 0.20${\rm R_{\sun}}$, in better agreement with the asteroseismology results. 

\begin{figure}[t]
\centering
\includegraphics[scale=0.45]{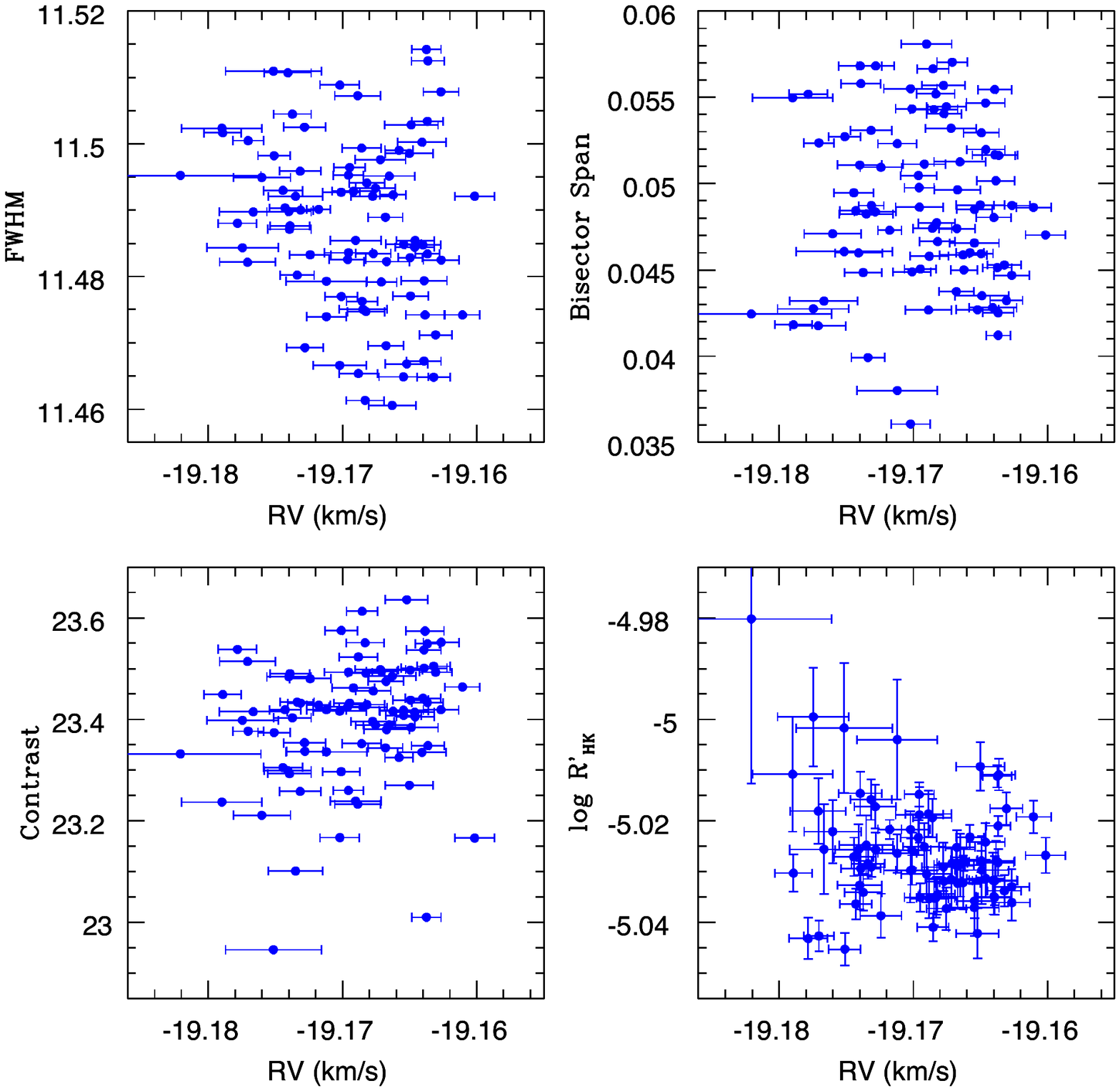}
\caption{Plots of the $\rm R^{'}_{HK}$ activity index and the FWHM, bisector velocity span, and contrast of the  cross-correlation function versus the radial velocity measurements from the HARPS-N data. There are no apparent correlations.\label{fig:ActivityIndexes}}
\end{figure}

Our analysis of the HARPS-N spectra yields a projected rotational velocity of $v\,\sin\,i_\star$ = 8.4  $\pm$ 0.5 ${\rm km~s^{-1}}$. The errorbars in the reported rotational velocity include uncertainties due to the line broadening by the spectrograph. From the HARPS-N spectra we also computed the $\rm R'_{HK}$ activity index and several parameters of the cross-correlation function (CCF), i.e. the FWHM, the Bisector span, and the Contrast, in search for correlations with the RVs. We find no correlation between the RVs and any of those parameters, as illustrated in Figure~\ref{fig:ActivityIndexes}. In addition, we derived the age and rotation period of the star following \citet{Mamajek2008}. We estimate an age for the star of 3.03 $\pm$ 0.35 Gyr, which agrees with the age of 2.84 $\pm$ 0.35 Gyr derived from asteroseismology \citep{Howell2012}, a rotational period of ${\rm P_{rot}}$ = 14.83  $\pm$ 2.41 days, and a ${\rm \langle log~R'_{HK}\rangle}$ = -5.027 $\pm$ 0.011. 

\begin{figure}[t]
\centering
\includegraphics[scale=0.45]{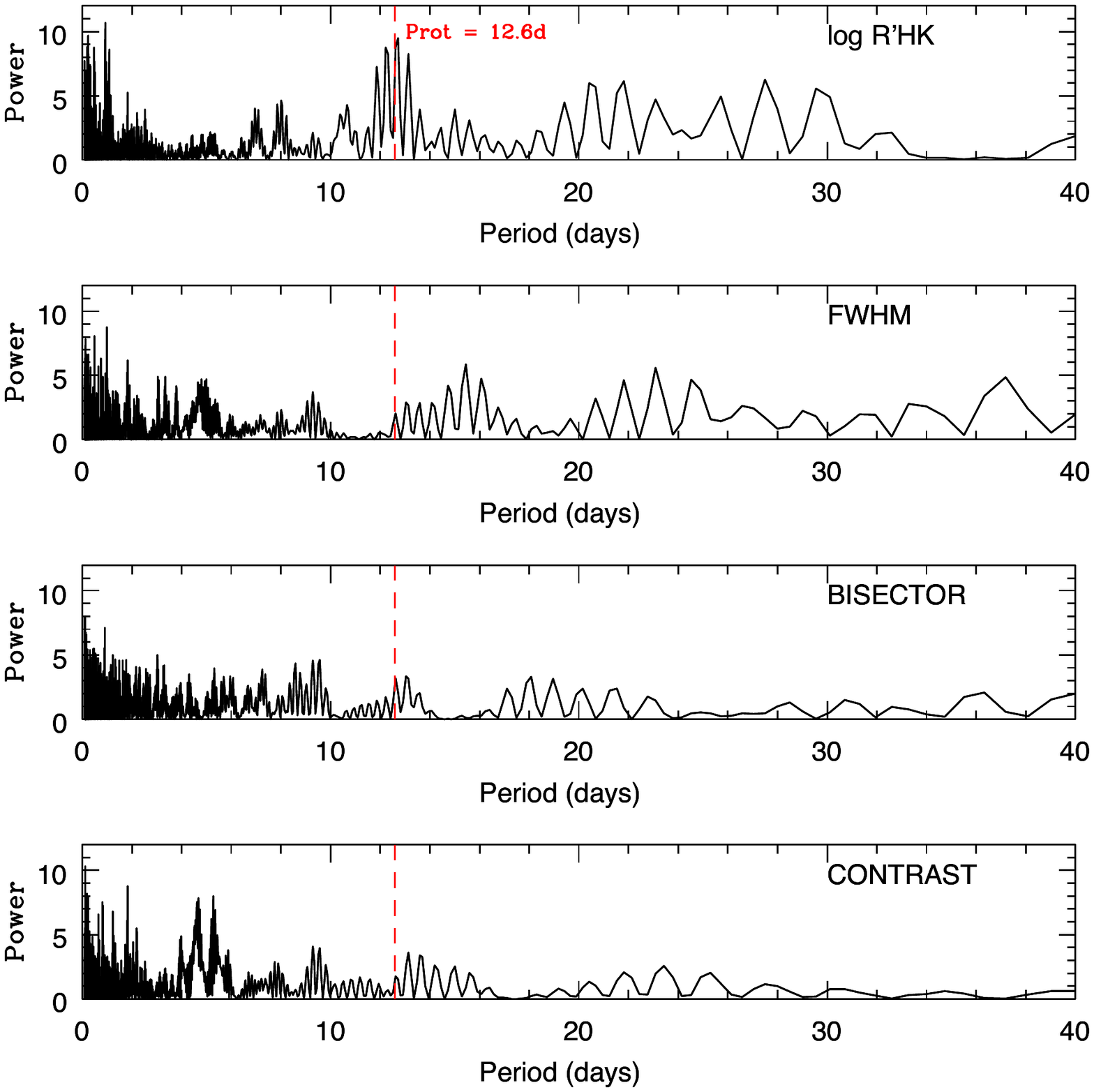}
\caption{Generalized Lomb-Scargle periodograms of the $\rm log~R'_{HK}$ activity index, FWHM,  Bisector span, and Contrast values obtained from the HARPS-N spectra. The  $\rm R'_{HK}$ activity index periodogram shows a clear peak at 12.6 days. \label{fig:RHKPeriodogram}}
\end{figure}

As a note, we also performed an GLS periodogram analysis of the $\rm R'_{HK}$ activity index values obtained from the HARPS-N spectra, as well as the FWHM,  Bisector span, and Contrast values and find a clear peak in the periodogram of the $\rm R'_{HK}$ index at 12.67 days (see Figure~\ref{fig:RHKPeriodogram}). This value coincides with the period of 12.6 days found by the ACF analysis of the Kepler LCs in section 3.1 and therefore reinforces the conclusion that 12.6 days corresponds to the rotation period of the star. The FWHM,  Bisector span, and Contrast periodograms do not show any strong peaks around that period.

\section{Light Curve and Radial Velocity Fits}

\subsection{Kepler Light Curve}

\begin{figure}[t]
\includegraphics[scale=0.45]{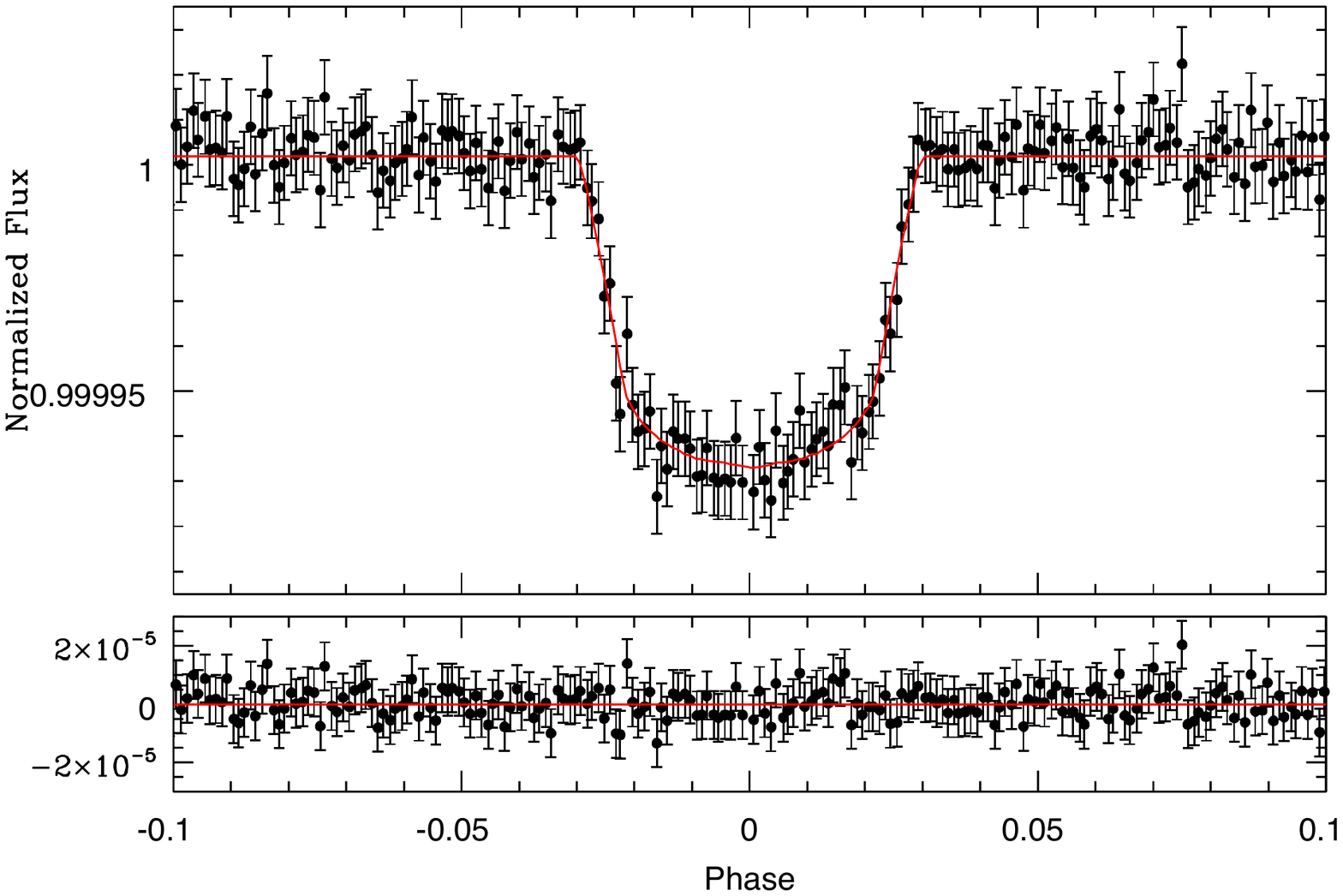}
\caption{{\it Top} -- Normalized DV light curve, phased using the planet period reported in Table 1. The best-fit transit model is shown in red. The black dots show the light curve data binned by factor of 1000, including the corresponding errorbars. {\it Bottom} -- Residuals to the transit fit. \label{fig:lcfit}}
\end{figure}

We fit the transit of Kepler-21b using the detrended Kepler-21b DV light curve shown in section ~\ref{sec:kepler} and EXOFAST \citep{Eastman2013}. In the fit, which only uses the long-cadence observations, we imposed Gaussian priors for the stellar parameters  ${\rm T_{eff}}$ = 6305 $\pm$ 50 K, ${\rm [Fe/H]}$ = -0.03 $\pm$ 0.10, and ${\rm log(g)}$ = 4.026 $\pm$ 0.004, based on the values reported by \cite{SilvaAguirre2015}.  We consider the parameters derived from asteroseismology more robust than those derived from the spectra, for the reasons explained in section 3.3. We also introduced a Gaussian prior for the normalized light curve baseline flux of ${\rm F_0}$ = 1.0000024 $\pm$ 0.00000031, to avoid systematic biases in the determination of the baseline flux introduced by in-transit points. We computed that normalized baseline flux beforehand by calculating the average and standard deviation of the light curve, not including the in-transit points and any other 3$\sigma$ outliers. The parameters fit for are the orbital period ${\rm P}$, the transit epoch ${\rm T_C}$, the semi-major axis to stellar radius ratio ${\rm a/R_*}$, the planet-to-star radius ratio ${\rm Rp/R_*}$, and the impact parameter ${\rm b = a/R_* ~cos~i}$, where ${\rm i}$ is the  orbital inclination. We used a quadratic limb darkening law, where the coefficients were not explicitly fit, but instead derived by interpolating the values in the \cite{Claret2011} tables for each value of ${\rm logg}$, ${\rm T_{eff}}$, and ${\rm [Fe/H]}$ in the fits. We modeled the system allowing for a non-zero eccentricity for Kepler-21b, but found solutions consistent with a circular orbit, also consistent with the analysis of the RVs. The results of the final fit, assuming a circular orbit, are summarized in Table~\ref{tab:lcfit}, which also includes a series of other parameters of the system computed by EXOFAST, e.g. the incident stellar flux in the surface of the planet, the transit probability, and the secondary eclipse time. The fit to the transit is also illustrated in Figure~\ref{fig:lcfit}. The parameter uncertainties in the table are derived using a Differential Evolution Markov Chain Monte Carlo method, as described in detail in section 2.2 of \cite{Eastman2013}.

\begin{deluxetable}{lcc}
\tabletypesize{\scriptsize}
\tablecaption{Median values and 68\% confidence interval for Kepler-21b.}
\centering
\tablehead{\colhead{~~~Parameter} & \colhead{Units} & \colhead{Value}}
\startdata
\sidehead{Stellar Parameters\footnotemark{}:}
                           ~~~$M_{}$\dotfill &Mass ($M_{\odot}$)\dotfill & $1.408_{-0.030}^{+0.021}$\\
                         ~~~$R_{}$\dotfill &Radius ($R_{\odot}$)\dotfill & $1.902_{-0.012}^{+0.018}$\\
                     ~~~$L_{}$\dotfill &Luminosity ($L_{\odot}$)\dotfill & $5.188_{-0.128}^{+0.142}$\\
                         ~~~$\rho_*$\dotfill &Density (cgs)\dotfill & $0.287_{-0.005}^{+0.004}$\\
              ~~~$\log(g_*)$\dotfill &Surface gravity (cgs)\dotfill & $4.026\pm0.004$\\
              ~~~$T_{eff}$\dotfill &Effective temperature (K)\dotfill & $6305\pm50$\\
                              ~~~$[Fe/H]$\dotfill &Metallicity\dotfill & $-0.03\pm0.10$\\
\sidehead{Planetary Parameters:}
                              ~~~$P$\dotfill &Period (days)\dotfill & $2.7858212\pm0.0000032$\\
                       ~~~$a$\dotfill &Semi-major axis (AU)\dotfill & $0.04285_{-0.00068}^{+0.00075}$\\
                           ~~~$R_{P}$\dotfill &Radius ($\rm R_{\oplus}$)\dotfill & $1.639_{-0.015}^{+0.019}$\\
           ~~~$T_{eq}$\dotfill & Equilibrium Temperature (K)\dotfill & $2025\pm20$\\
               ~~~$\langle F \rangle$\dotfill &Incident flux (${\rm 10^9}$ ${\rm erg}$ ${\rm s^{-1}}$ ${\rm cm^{-2}}$)\dotfill & $3.84\pm0.14$\\
\sidehead{Primary Transit Parameters:}
                ~~~$T_C$\dotfill &Time of transit ($BJD_{TBD}$)\dotfill & $2455093.83716_{-0.00085}^{+0.00082}$\\
~~~$R_{P}/R_{}$\dotfill &Radius of planet in stellar radii\dotfill & $0.007885\pm0.000050$\\
     ~~~$a/R_{}$\dotfill &Semi-major axis in stellar radii\dotfill & $4.929_{-0.047}^{+0.048}$\\
              ~~~$u_1$\dotfill &linear limb-darkening coeff\dotfill & $0.303_{-0.043}^{+0.044}$\\
           ~~~$u_2$\dotfill &quadratic limb-darkening coeff\dotfill & $0.296_{-0.046}^{+0.047}$\\
                      ~~~$i$\dotfill &Inclination (degrees)\dotfill & $83.20_{-0.26}^{+0.28}$\\
                           ~~~$b$\dotfill &Impact Parameter\dotfill & $0.584_{-0.020}^{+0.018}$\\
                         ~~~$\delta$\dotfill &Transit depth\dotfill & $0.00006217\pm0.00000078$\\
                ~~~$T_{FWHM}$\dotfill &FWHM duration (days)\dotfill & $0.1478_{-0.0018}^{+0.0019}$\\
          ~~~$\tau$\dotfill &Ingress/egress duration (days)\dotfill & $0.001784_{-0.000050}^{+0.000051}$\\
                 ~~~$T_{14}$\dotfill &Total duration (days)\dotfill & $0.1496\pm0.0019$\\
      ~~~$P_{T}$\dotfill &A priori non-grazing transit probability\dotfill & $0.2013_{-0.0019}^{+0.0020}$\\
                ~~~$P_{T,G}$\dotfill &A priori transit probability\dotfill & $0.2045\pm0.0020$\\
                            ~~~$F_0$\dotfill &Baseline flux\dotfill & $1.00000223_{-0.00000016}^{+0.00000015}$\\
\sidehead{Secondary Eclipse Parameters:}
              ~~~$T_{S}$\dotfill &Time of eclipse ($BJD_{TBD}$)\dotfill & $2455095.23007_{-0.00085}^{+0.00082}$\\
\enddata
\footnotetext[3]{values adopted from \cite{SilvaAguirre2015}} 
\label{tab:lcfit}
\end{deluxetable}

\subsection{Radial Velocities}

The RV, photometric, and  $\rm log~R'_{HK}$ activity index timeseries of Kepler-21 all show clear modulation around the stellar rotation period, as illustrated in figures~\ref{fig:LCPeriodogram},~\ref{fig:RVsPeriodograms}, and~\ref{fig:RHKPeriodogram}, which suggest the star is moderately active. Stellar activity hinders the detectability of the planetary signals in RV curves, but 
recent studies \citep{Haywood2014,Grunblatt2015,Rajpaul2015,Faria2016}, have succeeded on modeling the activity and extracted the planetary doppler signals using Gaussian Process regression \citep[hereafter referred to as GPs; see][for more details]{Rasmussen2006}. In an effort to extract the RV signal of Kepler-21b from the available data, we model the orbit of Kepler-21b as a Keplerian with free eccentricity, and model the correlated noise introduced by rotation-modulated stellar activity using a GP with a quasi-periodic covariance Kernel of the form
\begin{equation}
\label{cov}
k (t, t') = 
\eta_{1}^{2} \, \cdot \, 
\exp \left[ - \frac{(t - t')^{2}}{\eta_{2}^{2}}  - \frac{\sin^{2} \left( \frac{\pi (t - t')}{ \eta_{3}} \right)}{\eta_{4}^{2}}  \right] , 
\end{equation}

\noindent where the hyper-parameter $\eta_1$ is the amplitude of the covariance function, $\eta_2$ is equivalent to the evolution timescale of features in the stellar surface that produce activity-induced RV variations,  $\eta_3$ is equivalent to the stellar rotation period, and $\eta_4$ gives a measure of the level of high-frequency variability structure in the GP model. Our approach is similar to the ones used by \cite{Haywood2014}, \cite{Grunblatt2015} and \cite{Faria2016}, except that we set all the hyper-parameters as free parameters in the RV model. We leave $\eta_1$ as a free parameter, only constrained with a modified Jeffreys prior, as listed in Table~\ref{tab:priors}.  $\eta_2$ and $\eta_3$ are constrained with Gaussian priors using the values for the stellar rotation period and the active regions lifetime (or spot decay time) determined via the ACF analysis described in section 3.1. We constrain $\eta_4$ with a Gaussian prior centered around 0.5 $\pm$ 0.05. This value, which is adopted based on experience from previous datasets, allows the RV curve to have up to two or three maxima and minima, as is typical of stellar light curves and RV curves. Foreshortening of spots and other stellar surface features at the limb, and stellar limb darkening smooth stellar photometric and RV variations, which means that a curve with more structure than the allowed by this value of $\eta_4$ would be unphysical. The strong constraints on the hyper-parameters, particularly $\eta_4$, are ultimately incorporated into the model likelihood and provide a realistic fit to the activity-induced variations, as shown in Figures~\ref{fig:model} and~\ref{fig:phase}. We note that GP is not only robust, but also extremely flexible. Our aim in this analysis is not to test how well an unconstrained GP can fit the data, but to use all the prior knowledge on the system to model the activity-driven signal as best as possible.

We introduce in the model jitter terms for each instrument dataset. Kepler-21 is slightly evolved, thus its photosphere should have fewer and larger granules than main-sequence solar type stars \citep{Schwarzschild1975,Antia1984,Mathur2011}, and we expect larger activity signals in both the photometry and RVs during turnover convective timescales of a few hours. This is confirmed in both figures~\ref{fig:sampleLCs} and~\ref{fig:allrvs}. Therefore, we expect uncorrelated noise in a few-hours timescale to be a combination of both instrument systematics and residual granulation and stellar oscillation motions. Regarding HARPS-N, recent observations of the Sun \citep[][Phillips et al. in prep]{Dumusque2015} have shown the instrument has a random day-to-day offset with an rms of 0.9 ${\rm m~s^{-1}}$  (Phillips, private comm.). We account for this instrumental systematic by adding a noise term ${\rm \sigma_{harpsn,instr}}$ = 0.9 $\pm$ 0.1 ${\rm m~s^{-1}}$ in quadrature to the measured RV errorbars. This value of ${\rm \sigma_{harpsn,instr}}$ is constrained by a Gaussian prior. To estimate the granulation (and oscillation)-induced noise, we compute the inverse-variance weighted standard deviation of the residuals within each night, at each MCMC step, after subtracting the planet model. We note that this is only possible for the 2015 HARPS-N dataset, which has two observations per night, separated by a few hours. This noise term, ${\rm \sigma_{harpsn,gran}}$, is then added in quadrature to the measured RV errorbars, together with ${\rm \sigma_{harpsn,instr}}$. In the case of HIRES, they also collected several RV measurements per night, but those were collected consecutively, so they cannot be used to probe granulation over several hours timescales. We also do not have additional information about the intrinsic instrumental systematics of HIRES, so we use an overall free term, ${\rm \sigma_{hires}}$, to account for both instrument and granulation noise.

We also adopt Gaussian priors for the orbital period of the planet and the phase of the transits in the Keplerian fit, using the best fit values for those parameters computed in section 4.1. Finally, we account for instrumental zero-point offsets of the two spectrographs with two separate terms, ${\rm RV_{0,hires}}$, for HIRES and ${\rm RV_{0,harpsn}}$ for HARPS-N. We summarize the priors used for each free parameter of our RV model in Table~\ref{tab:priors}.

We fit the HIRES and HARPS-N data both separately and together. We maximize the likelihood of our model and determine the best-fit parameter values through a MCMC procedure similar to the one described in  \cite{Haywood2014}. We ran the MCMC chains for 1,000,000 steps each, confirming their convergence using the Gelman-Rubin criterion \citep{Gelman2004,Ford2006}. The best-fit parameters for all the three runs are summarized in Table~\ref{tab:bigtable}.

\subsubsection{HIRES-only analysis}

\begin{figure}
\includegraphics[scale=0.25]{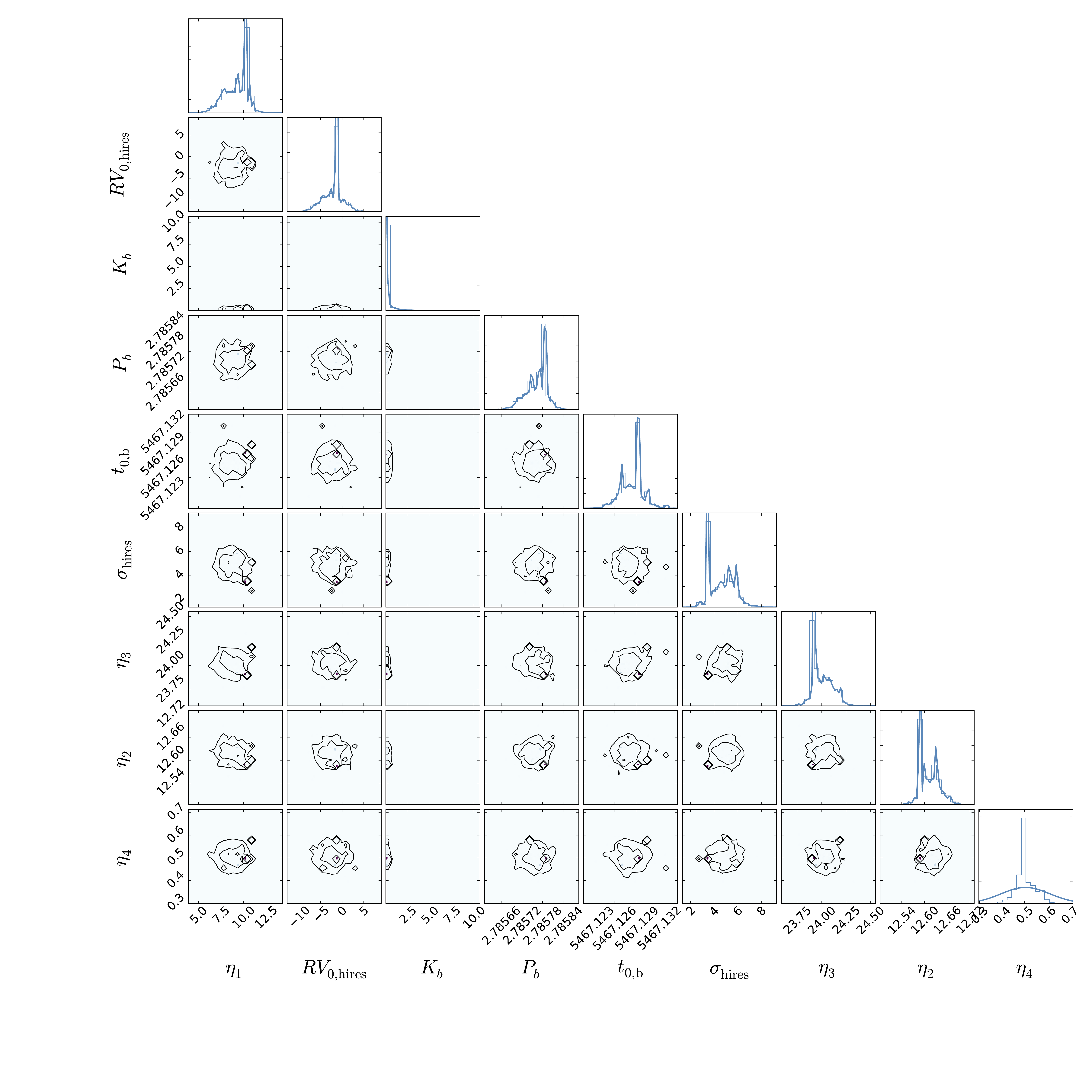}
\caption{Marginalized 1-D and 2-D posterior distributions of the model parameters when fitting the HIRES campaign only. 
The solid lines overplotted on the histograms are kernel density estimations of the marginal distributions.
It is clear that the MCMC chain has not converged, despite the additional priors imposed on $\eta_1$, $\rm RV_{0,hires}$, and ${\rm \sigma_{hires}}$ (the eccentricity was also kept fixed to zero).\label{badshark}}
\end{figure}

The HIRES RV dataset on its own yields no significant detection of Kepler-21b, as seen in column 1 of Table~\ref{tab:bigtable}. Our MCMC chains did not converge within 1,000,000 steps. As a test to see whether the chain converged using a more tightly constrained model, we fixed the eccentricity to zero and imposed Gaussian priors on the parameters that had uninformative priors otherwise: $\eta_1$, $\rm RV_{0,hires}$ and ${\rm \sigma_{hires}}$, but not ${\rm K_b}$. We centered these priors around the best-fit values of the combined HIRES and HARPS-N model. In spite of these constraints, the MCMC chains still do not converge as illustrated in Figure~\ref{badshark}. These findings are consistent with the non-detection result reported by \cite{Howell2012}. We also note that our best-fit values for ${\rm \sigma_{hires}}$ are consistent with the jitter value of 5 ${\rm m~s^{-1}}$ reported by \cite{Howell2012}.

We attribute the non-detection in the HIRES data to the adopted observing strategy. The HIRES dataset consists of 40 observations over 80 days. The first 39 observations consisted of groups of three consecutive 150s exposures, 
some collected on consecutive nights, and some with several-night gaps between them. Each group of three consecutive observations was averaged out to produce an RV data point per night, equivalent to a 10 minute-long observation, 
so the 40 observations dataset is effectively equivalent to 14 observations. The largest stretch of consecutive nightly observations is five days, which is slightly less than two orbital periods of Kepler-21b.  The full run spans 80 nights, which 
based on our analysis in section 3.1, span three stellar active-region evolution timescales, and over six times the stellar rotation cycle. Therefore, the coherence in the activity-induced signal is lost and the sampling of the observations is too
sparse to decouple the orbit of Kepler-21b from the activity-induced stellar variations.

\subsubsection{HARPS-N-only analysis}

The HARPS-N dataset, although its observing strategy was not fully tailored for this system either, does yield a detection of the doppler signal of Kepler-21b and a mass measurement of 5.41 $\pm$ 1.76 $\rm M_{\oplus}$ (see column 2 in Table~\ref{tab:bigtable}). 
The HARPS-N observations are split into two seasons. During the 2014 season, of 45 observations spanning almost four active-region lifetimes and eight stellar rotation cycles, we collected a single observation per night, with the exception of one night in the middle of the run and the last four nights of the run, when we collected two observations per night. Two thirds of the observations were taken in three clusters of 6, 7, and 11 consecutive nights; the remaining third are scattered in groups of three nights or fewer. In the 2015 season we deployed a different observing strategy, which proved to be better suited to characterizing the planet's orbit and the activity signal. In this season we collected 37 observations: the first five were still one observation per night, on isolated nights; the other 32 observations were taken twice per night, separated always at least by 3.5 hours, but on average five hours. The last 28 observations were taken on 15 consecutive nights, with gaps in two of the nights due to bad weather. This last stretch of observations fits well within one active-region lifetime (about two stellar rotation cycles), which means that the coherency of the activity-induced variations is preserved. However, we note that the 28 last observations alone do not yield a significant detection of Kepler-21b. The ideal number of observations, their sampling and stretch in time must be somewhere between these 28 observations and the full HARPS-N campaign, but exploring this further is beyond the scope of our analysis in this paper. However, we highlight that observational strategies customized for individual targets will yield to significant improvement in our capacity to detect planetary doppler signals in RV curves of active stars.

\subsubsection{HIRES and HARPS-N combined analysis}

\begin{figure}[t]
\centering
\includegraphics[scale=0.6, angle=90]{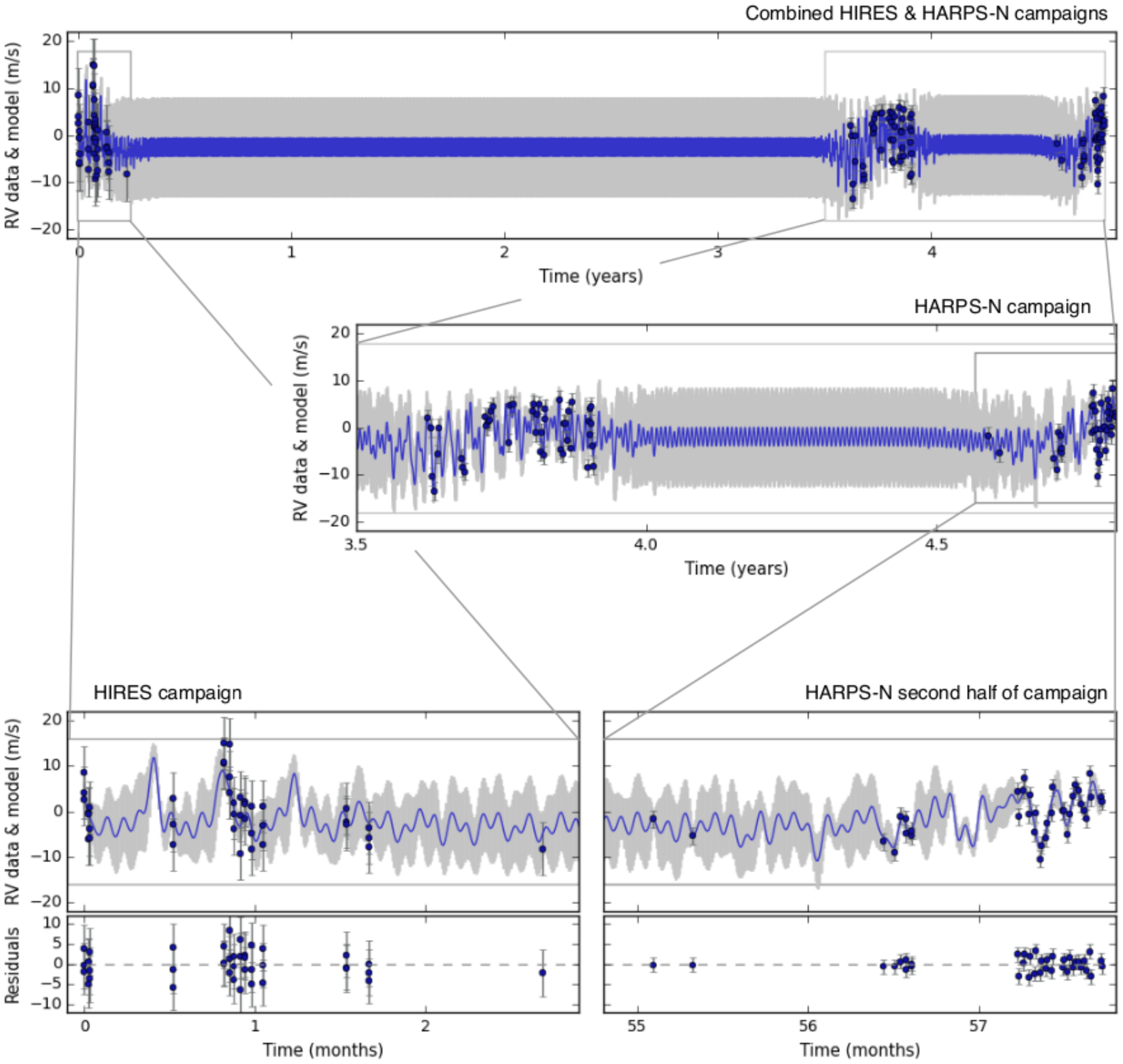}
\caption{The HIRES and HARPS-N RV data (black points with error bars) and our best fit (blue line with grey shaded 1-$\sigma$ error regions), over various timescales: \emph{(top)} the full combined dataset spanning approximately 5 years -- the first campaign was obtained with HIRES, while the second one is from HARPS-N; \emph{(middle)} zoom-in on the HARPS-N dataset spanning just under 1.5 years; \emph{(bottom left)} zoom-in on the 3-month HIRES dataset; \emph{(bottom right)} zoom-in on a portion of HARPS-N data also covering 3 months. The two bottom panels, of equal timespan, show how the frequency structure of the model is preserved in time and adequately fits the observations throughout the combined dataset. The residuals after subtracting the model from the data are shown for the two bottom panels (they are representative of the whole dataset). \label{fig:model}}
\end{figure}

\begin{figure}[t]
\centering
\includegraphics[scale=0.7, angle=90]{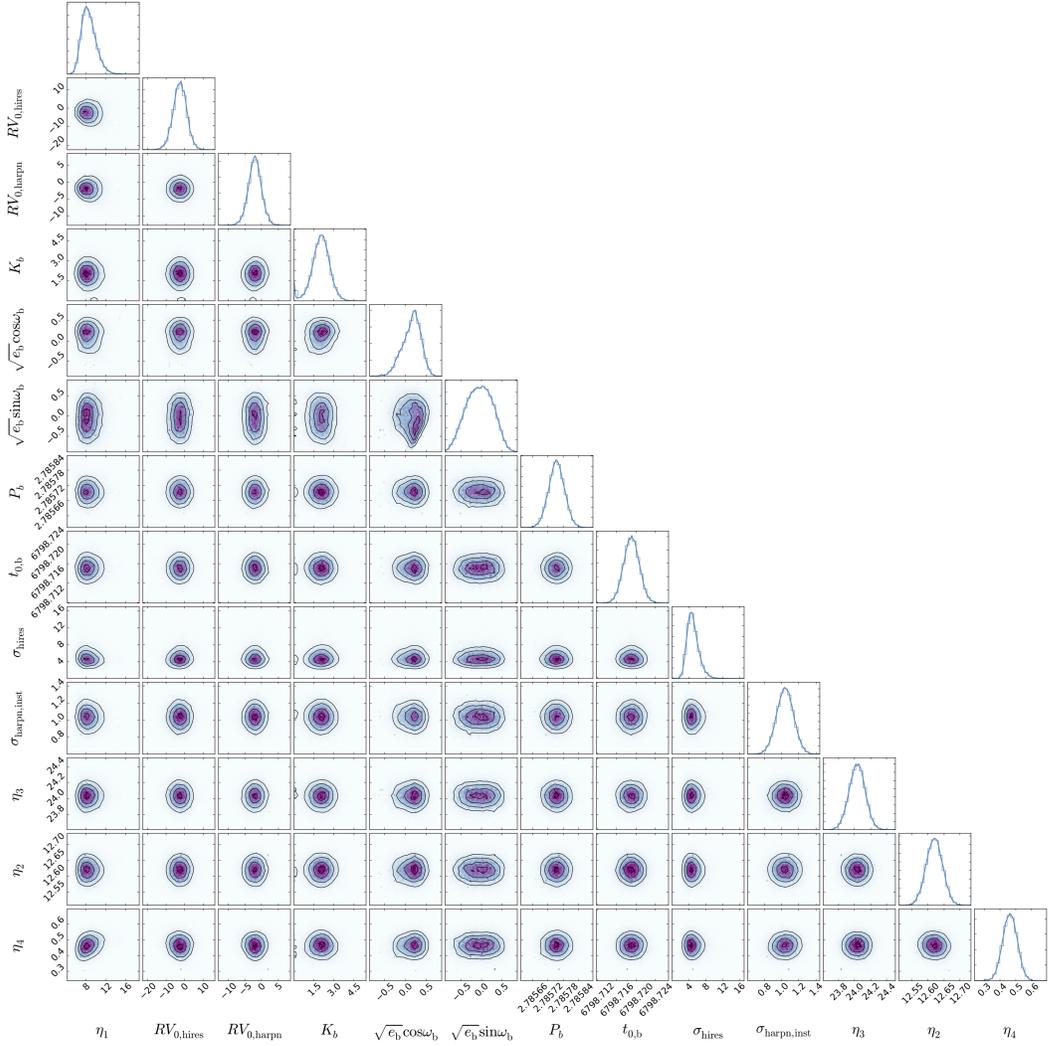}
\caption{Marginalized 1-D and 2-D posterior distributions of the model parameters when fitting the HIRES and HARPS-N campaigns together. 
The solid lines overplotted on the histograms are kernel density estimations of the marginal distributions.
These smooth, Gaussian-shaped posterior distributions attest of the good convergence of the MCMC chain.\label{goodshark}}
\end{figure}

Although the HIRES campaign alone does not contain sufficient information to provide a robust mass determination of Kepler-21b, those data are still compatible with the HARPS-N campaign, as illustrated in Figure~\ref{fig:model}, and their combined analysis yields system parameters fully consistent with the HARPS-N data alone, as shown in column 3 of Table~\ref{tab:bigtable}. Figure~\ref{goodshark} shows the posterior distributions of the MCMC analysis for the combined datasets. Looking closely at the correlation plots of ${\rm K_b}$ reveals, in addition to the main distribution peak, a local area of maximum likelihood near ${\rm K_b}$ = 0 ${\rm m~s^{-1}}$, which is not present in the analysis of the HARPS-N data alone and is therefore introduced by the HIRES data.  This may be interpreted as the HIRES data acting as a prior on the HARPS-N observations when we combine the two datasets. In this case, the influence of this peak is diminished by the larger sampling of the HARPS-N data and the posterior distribution of ${\rm K_b}$ is fully dominated by the HARPS-N dataset (as shown in Table~\ref{tab:bigtable} this does not significantly affect the resultant planet mass). However, there may be cases in which the additional peak in the posterior distribution becomes more prominent, affecting the final fit. This could for example occur when combining datasets from different instruments with similar numbers of observations. 

\begin{figure}[t]
\centering
\includegraphics[scale=0.5]{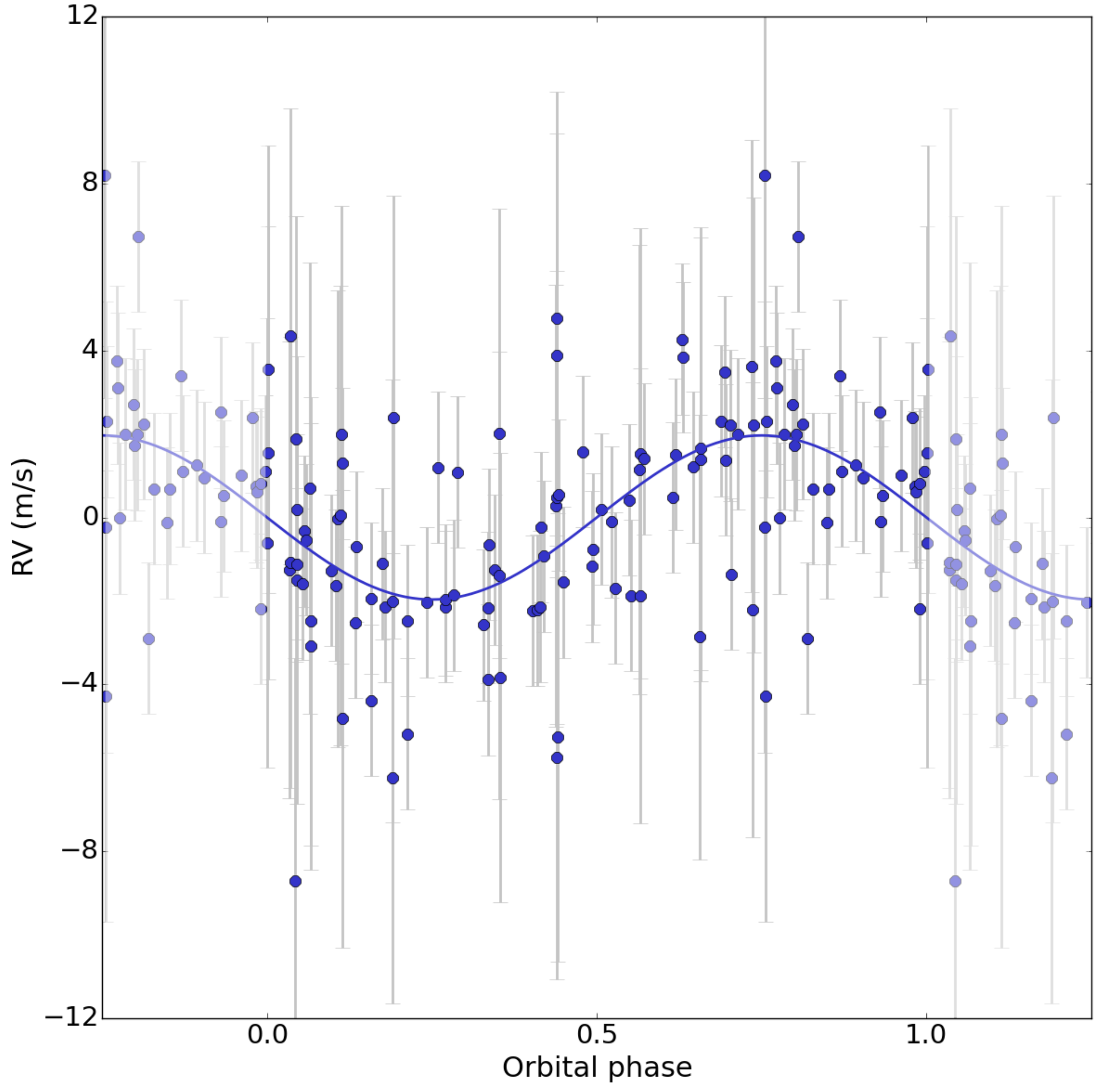}
\caption{Phase plot of the orbit of Kepler-21b for the best-fit model after subtracting the Gaussian process component, for the combined HIRES and HARPS-N datasets. \label{fig:phase}}
\end{figure}

Given that the results from the analysis of the HARPS-N data alone and the combined datasets are consistent, we adopt the solution of the combined datasets (column 3 Table~\ref{tab:bigtable}) in as the best-fit parameter values for the system. The resulting best-fit model for the combined dataset is shown in Figure~\ref{fig:model}, and the resultant phase folded orbit of Kepler-21b is shown in Figure~\ref{fig:phase}. We note that the value of ${\rm K_b}$ for each fit remains consistent withing 1$\sigma$ regardless of our choice of covariance function, parameter distributions or initial parameter values, which attests the robustness of this result. In addition, the best-fit parameters remain the same when we fix the orbital eccentricity to zero, which supports the result of a zero eccentricity measurement in Table~\ref{tab:bigtable}.

\begin{table}
\caption{\footnotesize{Best-fit parameter values of the RV model for the HIRES-only, HARPS-N-only and combined datasets. The numbers in parentheses represent the uncertainty in the last digit of the value. }}
\flushleft
\centering
\scriptsize
\begin{minipage}{120mm}

\begin{tabular}{llllllll}

\hline
\hline
\rule{0pt}{4ex} & HIRES only  & HARPS-N only & HIRES + HARPS-N  \\  

\hline
\multicolumn{1}{l}{\rule{0pt}{4ex}Kepler-21b} \\
\rule{0pt}{0ex} \\
$P$ [days] 				& $2.78578 (3) $
						& $2.78574 (3) $
						& $2.78578 (3) $ \\				

$t_{0,{\rm b}}$ [BJD - 2450000] & $2456798.7188 (1)  $
						& $2456798.7188 (1)  $
						& $2456798.7188 (1)  $ \\	

$t_{\rm peri, b}$ [BJD - 2450000] & $2456798.2  \pm 0.7$
						& $2456798.2  \pm 0.7$
						& $2456797.9  \pm 0.7$ \\

$K_{\rm b}$ [m~s$^{-1}$] 			& $0.3 \pm 1$
							& $2.12 \pm 0.66$
							& $1.99\pm 0.65$ \\
			
$m_{\rm b}$ [M$_{\oplus}$]		& $0.7 \pm 2.5$
							& $ 5.41\pm 1.76$
							& $5.08\pm 1.72$ \\
							
$e_{\rm b}$ 					& $ 0.006\pm 0.2$
							& $ 0.007\pm 0.1$
							& $0.02 \pm 0.1$\\

$\omega_{\rm b}$ [$^\circ$]  		& $ -106\pm104 $
							& $2 \pm 89$
							& $ -15\pm 79$ \\

$a_{\rm b}$ [AU] 				& $ 0.0427172 (3)$
							& $0.0427172 (3) $
							& $0.0427172 (3)$ \\
				
\rule{0pt}{0ex} \\

\hline
\multicolumn{3}{l}{\rule{0pt}{4ex}Hyper-parameters of the GP} \\
\rule{0pt}{0ex} \\

$\eta_1$  [m~s$^{-1}$] 			& $ 8.9\pm 6.6$
							& $ 6.7\pm 1.4$
							& $ 8.6\pm 1.4$ \\

$\eta_2$  [days] 				& $ 24.21\pm 0.1$
							& $ 24.04\pm0.09 $
							& $ 23.95\pm 0.09$ \\

$\eta_3$  [days] 				& $ 12.61\pm0.02 $
							& $ 12.60\pm 0.02$
							& $12.63 \pm 0.02$ \\

$\eta_4$						& $ 0.50\pm0.05 $
							& $ 0.42\pm 0.05$
							& $ 0.45\pm 0.05$ \\

\rule{0pt}{0ex} \\

\hline
\multicolumn{3}{l}{\rule{0pt}{4ex}Uncorrelated noise terms} \\
\rule{0pt}{0ex} \\

$\sigma_{\rm hires}$	 [m~s$^{-1}$] 			& $ 5.4\pm 1.5$
							& --
							& $ 4.9\pm 1.4$ \\

$\sigma_{\rm harpsn, instr}$ [m~s$^{-1}$] 			& --
							& $ 0.9\pm 0.1$
							& $ 1.0\pm0.1 $ \\
							
$\sigma_{\rm harpsn, gran}$ [m~s$^{-1}$] 		& --
							& $1.73 \pm 0.04$
							& $ 1.50\pm 0.03$ \\							
\rule{0pt}{0ex} \\

\hline
\multicolumn{3}{l}{\rule{0pt}{4ex}Systematic RV offsets} \\
\rule{0pt}{0ex} \\

$RV_{0, {\rm hires}}$ [m~s$^{-1}$] 			& $ -2.2\pm4.7 $
							& --
							& $ -2.5\pm 3.8$ \\

$RV_{0, {\rm harpsn}}$ [m~s$^{-1}$] 			& --
							& $ -10.0\pm 1.6$
							& $ -10.5\pm 2.0$ \\
\rule{0pt}{0ex} \\
\hline
\label{tab:bigtable}
\end{tabular}

\end{minipage}
\end{table}

\begin{table}
\caption{\footnotesize{Parameters modeled in the RV analysis and their prior probability distributions.}}
\label{tab:priors}
\begin{threeparttable}
\hspace{10pt}
\footnotesize
\begin{tabular}{@{}llllllllllllll@{}}
\hline
\hline

\hline
\multicolumn{3}{c}{Kepler-21b orbital parameters} \\
\hline
\rule{0pt}{0ex} \\

$P_{\rm orb, b}$			&
Orbital period
&	Gaussian	($P_{\rm orb, b}, \sigma_{\rm P_{\rm orb, b}}$)			\\

$t_{\rm 0, b}$			&		
Transit ephemeris		
&	Gaussian	($t_{\rm 0, b}, \sigma_{\rm t_{\rm 0, b}}$)		\\
										
$K_b$					&		
RV semi-amplitude		
&	Modified Jeffreys ($\sigma_{RV}$, $2\,\sigma_{RV}$)				\\

$e_{\rm b}$			
& Orbital eccentricity		& 	Square root [0, 1]			\\

$\omega_{\rm b}$			 
& Argument of periastron	& 	Uniform [$0, 2\pi$]							\\

\rule{0pt}{0ex} \\
\hline
\multicolumn{3}{c}{Hyper-parameters of the GP} \\
\hline
\rule{0pt}{0ex} \\

$\eta_1$&
Amplitude of covariance 
& 	Modified Jeffreys ($\sigma_{RV}$, $2\,\sigma_{RV}$) 				\\

$\eta_2$&
Evolution timescale
&	Gaussian	($T_{\rm ev}, \sigma_{\rm T_{ev}}$)		\\

$\eta_3$&
Recurrence timescale
&	Gaussian	($P_{\rm rot}, \sigma_{\rm P_{rot}}$)		\\

$\eta_4$&
Structure parameter
&	Gaussian	(0.5, 0.05)		\\

\rule{0pt}{0ex} \\
\hline
\multicolumn{3}{c}{Uncorrelated noise terms} \\
\hline
\rule{0pt}{0ex} \\

$\sigma_{\rm hires}$	& HIRES instrument + granulation

& 	Jeffreys [0.01, 10 m~s$^{-1}$] 									\\	

$\sigma_{\rm harpsn, instr}$	& HARPS-N instrument

& 	Jeffreys [0.01, 10 m~s$^{-1}$] 									\\

$\sigma_{\rm harpsn, gran}$	& HARPS-N granulation

& 	Determined in MCMC	\\

\rule{0pt}{0ex} \\
\hline
\multicolumn{3}{c}{Systematic RV offsets} \\
\hline
\rule{0pt}{0ex} \\

$RV_{0, {\rm hires}}$	&
HIRES dataset
& 	Uniform 									\\

$RV_{0, {\rm harpsn}}$	&
HARPS-N dataset
& 	Uniform 									\\			
\hline

\vspace{20pt}
\end{tabular}
 \begin{tablenotes}
            \item[] {\bf Notes.} For modified Jeffreys priors, the terms in the brackets refer to the knee and maximum value of the prior. In the case of a Gaussian prior, the terms within brackets represent the mean and standard deviation of the distribution. The terms within square brackets stand for the lower and upper limit of the specified distribution; if no interval is given, no limits are placed.
        \end{tablenotes}
     \end{threeparttable}
\end{table}

\section{Discussion and Conclusions}

\begin{figure}[t]
\centering
\includegraphics[scale=0.45]{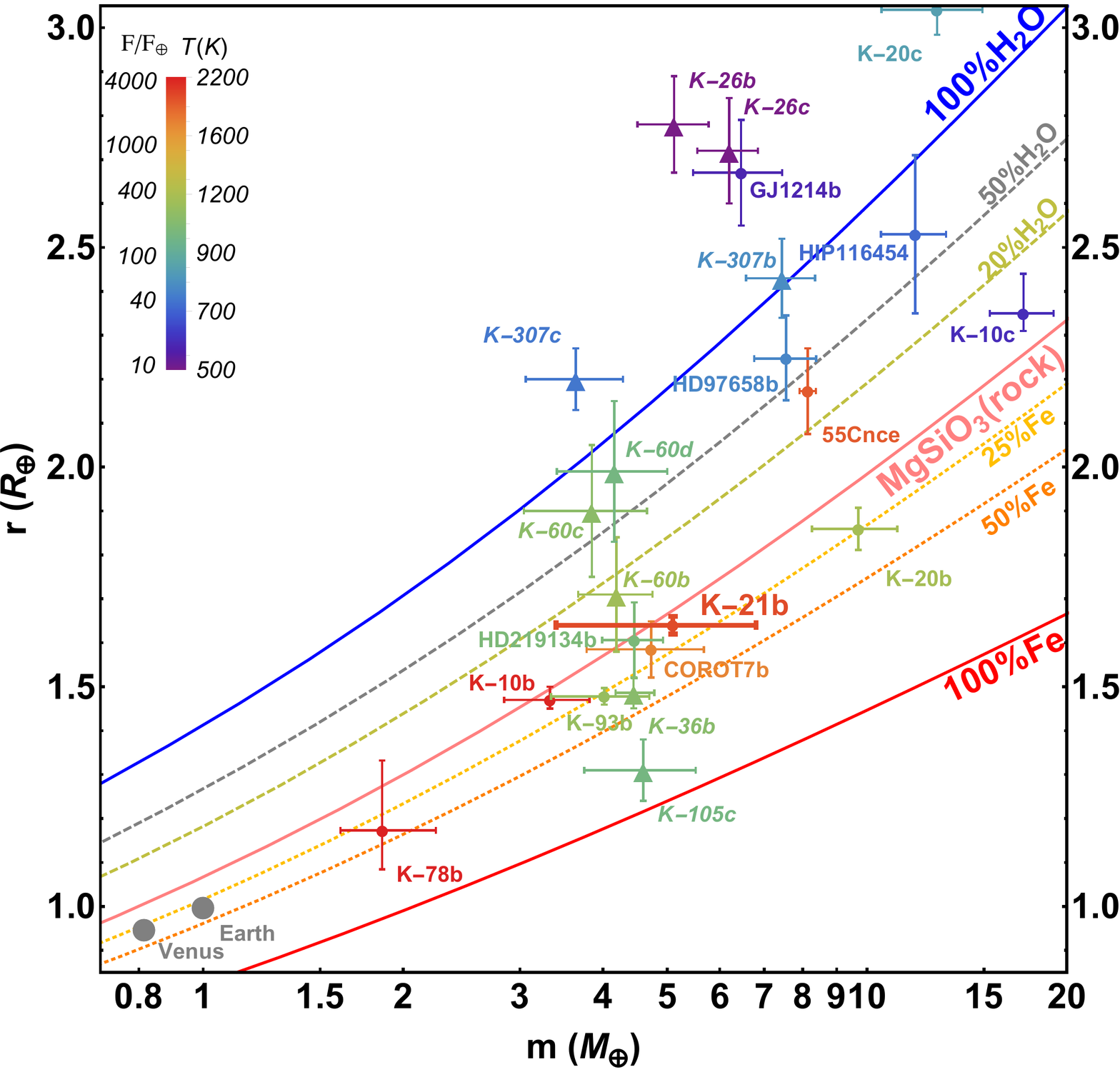}
\caption{Mass-Radius relation for planets with masses $<$ 20 $\rm M_{\oplus}$, measured with precisions better than 20$\%$. Circles indicate the planets with masses measured via RVs; triangles indicate planets with masses measured via TTVs \citep{Carter2012,JontofHutter2016}. The plot also includes Earth and Venus, for reference. The lines show models of different compositions, with solid lines indicating {\it single} composition planets (either $\rm H_2O$, $\rm MgSiO_3$, i.e. rock, or $\rm Fe$). The dashed and dotted lines indicate Mg-silicate planets with different amounts of $\rm H_2O$ and $\rm Fe$.  The data points representing the planets are color-coded as a function of incident bolometric stellar flux (compared to the Earth) and equilibrium temperature (assuming circular orbit, uniform planetary surface temperature, and bond albedo A=0). For other A values, the temperature can be obtained by multiplying those values by a factor of $\rm (1-A)^{1/4}$, following the flux and temperature scale indicated in the upper-left corner of the diagram.} \label{fig:massradius}
\end{figure}

We measure a mass for Kepler-21b of 5.1 $\pm$ 1.7 $\rm M_{\oplus}$ and derive a revised radius for the planet of 1.639$^{\rm +0.019}_{\rm -0.015}$ $\rm R_{\oplus}$, in agreement with the previous radius measurement of \cite{Howell2012}. Those parameters combined yield a density for this object of 6.4 $\pm$ 2.1  ${\rm g~cm^{-3}}$, which suggests a rocky composition. Figure~\ref{fig:massradius} shows theoretical mass-radius curves for planets composed of 100$\%$, 50$\%$, and 25$\%$ $\rm H_2O$, as well as rocky planets with 25$\%$, 50$\%$, and 100$\%$ $\rm Fe$ cores and the remaining mass in magnesium silicate mantles \citep{Zeng2016}. The figure also shows all the mass-radius measurements so far for exoplanets with masses less than 20 $M_{\oplus}$ and mass errors smaller than 20$\%$. The location of Kepler-21b in this diagram is consistent with a rocky composition. Kepler-21b fits within the group of 1--6 $\rm M_{\oplus}$ planets reported by \cite{Dressing2015b} as being well-described by the same fixed ratio of iron to magnesium silicate. The recently discovered Kepler-20b, with a mass of 9.7 $M_{\oplus}$ also fits in that group \citep{Buchhave2016}. Kepler-21b has also similar parameters to CoRoT-7b \citep{Barros2014,Haywood2014}.

If the interior of Kepler-21b is differentiated, i.e. the $\rm Fe$ in the planet's interior has sunk to the center, while the lighter silicates remain in the mantle, we can use eq.~3 in \cite{Zeng2016} to estimate a core mass fraction (CMF) for this planet of 0.1 $\pm$ 0.3, which is, within the uncertainties, close to the CMF of 0.3 for Earth and Venus in the Solar System. Most of the uncertainty in this CMF estimate comes from the current error in the mass and refining the mass measurement would yield a more accurate CMF estimate. Rocky planets of the same composition and the same mass, one differentiated, one un-differentiated, will have almost identical radius, within 1-2$\%$ \citep{Zeng2013}, so at present we cannot distinguish between these two scenarios given the current uncertainty in the radius of Kepler-21b of 1.2$\%$.

With an estimated equilibrium temperature of about 2000 K, the top few-hundred-kilometer-thick layer of Kepler-21b is expected to be molten. However, the silicate (rocky) mantle underneath is expected to be solid due to fact that the adiabat has shallower slope than the melting curve \citep{Zeng2016b,Stixrude2014}. The core of the planet is expected to be fully or partially molten. An interior structure calculation for Kepler-21b using the {\it Manipulate Planet} tool \citep{Zeng2016,Zeng2014,Zeng2013}, gives a central pressure for the planet of around 1200 ${\rm GPa}$. The pressure at the core-mantle boundary is estimated to be 800 ${\rm GPa}$. The density at the planet's center is estimated to be about 17 ${\rm g~cc^{-1}}$, so compared to the zero-pressure density of iron (7-8 ${\rm g~cc^{-1}}$), there appears to be significant compression in the core. The density of silicate at the core-mantle boundary of the planet is estimated to be about 8 ${\rm g~cc^{-1}}$.



Kepler-21b orbits the brightest planet host star discovered by the ${\it Kepler}$ mission. The star is a slightly evolved F6IV subgiant, with intrinsic radial velocity variations up to about 10 ${\rm m~s^{-1}}$. With Gaussian Process regression, however, we can reconstruct the intrinsic stellar variability well enough to confidently extract radial velocity signals with amplitudes five times smaller than the stellar noise. The apparent brightness of Kepler-21 is similar to the bright targets to be observed by TESS and many of those targets will most likely have significant intrinsic radial velocity variability. Therefore, this study serves as example of the kind of radial velocity analysis that will be necessary to confirm the masses of TESS planet candidates. In particular, we emphasize the need for radial velocity observations with cadence tailored for each target, based on their stellar rotational period and active-region lifetimes to efficiently model the activity and extract the planetary doppler signal.

Given the proximity of Kepler-21b to its host star and with a planetary surface temperature of about 2000K, it is unlikely that the planet has retained a significant amount of envelope volatiles. However, even though the atmosphere of the planet is expected to be tenuous, the brightness of the system may allow detection of atmospheric features in the UV, optical or infrared, either from space, with HST and JWST, or from the ground with upcoming large facilities.

\acknowledgments

The thank the anonymous referee for helpful comments to the manuscript. The HARPS-N project has been funded by the Prodex Program of the Swiss Space Office (SSO), the Harvard University Origins of Life Initiative (HUOLI), the Scottish Universities Physics Alliance (SUPA), the University of Geneva, the Smithsonian Astrophysical Observatory (SAO), and the Italian National Astrophysical Institute (INAF), the University of St. Andrews, Queens University Belfast, and the University of Edinburgh. This publication was made possible through the support of a grant from the John Templeton Foundation. The opinions expressed are those of the authors and do not necessarily reflect the views of the John Templeton Foundation. This material is based upon work supported by the National Aeronautics and Space Administration under grant No. NNX15AC90G issued through the Exoplanets Research Program. The research leading to these results has received funding from the European Union Seventh Framework Programme (FP7/2007-2013) under grant Agreement No. 313014 (ETAEARTH). This work was performed in part under contract with the California Institute of Technology (Caltech)/Jet Propulsion Laboratory (JPL) funded by NASA through the Sagan Fellowship Program executed by the NASA Exoplanet Science Institute. P.F. acknowledges support by Fundacao para a Ciencias e a Tecnologia (FCT) through Investigador FCT contract of reference IF/01037/2013 and POPH/FSE (EC) by FEDER funding through the program "Programa Operacional de Factores de Competitividade - COMPETE", and further support in the form of an exploratory project of reference IF/01037/2013CP1191/CT0001. L.Z. is supported by a grant from the Simons Foundation (SCOL, award $\#$337090).  A.V. is supported by the NSF Graduate Research Fellowship, Grant No. DGE 1144152.  X.D. is grateful to the Society in Science-Branco Weiss Fellowship for its financial support.

{\it Facilities:} \facility{TNG:HARPS-N}, \facility{Keck:HIRES}, \facility{TRES}.









\clearpage




\clearpage

\begin{deluxetable}{lrrrrlr}
\tabletypesize{\normalsize}
\tablecolumns{7}
\tablewidth{0pt}
\tablecaption{HARPS-N radial velocity data$^{a}$ \label{tab:data}}
\tablehead{ \colhead{$\rm BJD_{UTC}$}    & \colhead{RV}                    & \colhead{$\sigma_{\rm RV}$}          & \colhead{BIS$_{\rm span}$}          & \colhead{log $\rm R_{HK}^{'}$} &  \colhead{$\sigma$\rm log $R_{HK}^{'}$}  &   \colhead{$\rm t_{exp}$}\\
\colhead{$-2\,400\,000$}   & \colhead{(m s$^{-1}$)}      & \colhead{(m s$^{-1}$)}                     & \colhead{(m s$^{-1}$)}                  & \colhead{(dex)}                     & \colhead{(dex)}                                      & \colhead{(s)}
}
\vspace{10pt}
\startdata
56762.639368&-19166.52	&1.93&51.26&-5.0287&0.0062&1800\\
56764.647510&-19168.57	&1.45&47.41&-5.0194&0.0038&1800\\
56765.622661&-19178.93	&1.38&41.82&-5.0303&0.0037&1200\\

56766.637190&-19182.07	&5.99&42.44&-4.9802&0.0324&  900\\	
56768.669315&-19174.07	&1.75&45.98&-5.0257&0.0050&1800\\	
56769.737457&-19168.53	&1.16&56.65&-5.0410&0.0028&1200\\
56783.561093&-19175.10	&1.19&52.72&-5.0453&0.0032&1800\\
56784.564843&-19177.03	&1.15&52.36&-5.0427&0.0030&1800\\
56785.617658&-19177.83	&1.43&55.19&-5.0432&0.0041&1800\\
56798.576307&-19166.23	&0.99&45.00&-5.0277&0.0022&1800\\
56799.561359&-19168.19	&1.32&46.64&-5.0345&0.0034&  900\\	
56800.551758&-19167.10&1.14&57.04&-5.0283&0.0029&1200	\\
56801.524205&-19166.80	&1.30&43.74&-5.0288&0.0033&1200\\
56802.539813&-19164.89	&2.00&43.50&-5.0297&0.0067&1800\\
56803.572851&-19164.08	&1.79&42.81&-5.0317&0.0054&1800\\
56813.516317&-19171.77	&0.88&47.29&-5.0217&0.0019&1800\\
56814.537057&-19163.74	&1.07&45.13&-5.0112&0.0022&1800\\
56816.639846&-19163.67	&0.90&41.21&-5.0210&0.0020&1800\\
56828.455942&-19165.01	&1.78&48.77&-5.0093&0.0048&1800\\
56829.423431&-19163.63	&1.23&51.63&-5.0109&0.0031&1800\\
56830.454195&-19169.56	&1.11&49.77&-5.0148&0.0024&1800\\
56831.713969&-19165.78	&1.04&46.00&-5.0232&0.0024&1800\\
56832.468881&-19163.67	&1.15&42.51&-5.0282&0.0027&1800\\
56833.531472&-19173.73	&1.38&44.85&-5.0341&0.0036&1800\\
56834.451861&-19169.47	&1.17&45.05&-5.0351&0.0028&1800\\
56835.497761&-19174.28	&1.18&48.43&-5.0364&0.0030&1800\\
56836.452270&-19166.71	&1.71&49.64&-5.0323&0.0051&1800\\
56836.614171&-19164.59	&1.42&51.97&-5.0314&0.0037&1800\\
56845.433549&-19162.66	&1.31&44.69&-5.0330&0.0035&1800\\
56846.481384&-19173.12	&0.94&48.73&-5.0291&0.0020&1800\\
56847.439704&-19173.99	&1.64&51.08&-5.0327&0.0048&1800\\
56848.402826&-19167.54	&1.46&54.46&-5.0373&0.0042&1800\\
56849.404363&-19167.75	&1.61&55.69&-5.0291&0.0048&1800\\
56850.412033&-19171.20	&3.00&38.01&-5.0040&0.0118&1800\\
56851.408417&-19164.95	&0.91&45.94&-5.0280&0.0021&1800\\
56852.407578&-19172.83	&1.56&48.37&-5.0172&0.0043&1800\\
56853.411021&-19163.06	&1.22&43.22&-5.0176&0.0032&1800\\
56863.638019&-19177.08	&2.06&41.76&-5.0181&0.0065&1500\\
56863.680716&-19167.18	&1.88&53.21&-5.0315&0.0061& 600\\
56864.501604&-19170.09	&1.21&54.33&-5.0297&0.0029&1500\\
56864.622460&-19164.62	&1.43&54.67&-5.0242&0.0037&900\\
56865.515222&-19167.71	&1.30&54.05&-5.0319&0.0032&900\\
56865.651969&-19163.92	&1.69&51.67&-5.0318&0.0048&600\\
56866.489974&-19172.41	&1.54&50.95&-5.0387&0.0043&900\\
56866.602485&-19176.65	&2.53&43.19&-5.0256&0.0089&600\\
57115.684925&-19170.21	&1.97&55.50&-5.0298&0.0063&900\\
57122.653637&-19173.91	&1.47&55.80&-5.0294&0.0042&900\\
57156.575680&-19175.15	&3.58&46.06&-5.0017&0.0128&900\\
57158.600352&-19177.45	&2.66&42.74&-4.9995&0.0097&900\\
57159.594238&-19169.55	&1.78&48.63&-5.0188&0.0055&900\\
57160.591769&-19170.20	&1.46&36.07&-5.0217&0.0035&900\\
57160.701948&-19173.38	&1.25&39.92&-5.0284&0.0031&900\\
57161.578918&-19173.95	&1.62&56.82&-5.0146&0.0043&900\\
57161.695126&-19172.80	&1.37&56.83&-5.0257&0.0034&900\\
57180.506372&-19164.00	&1.32&48.04&-5.0349&0.0035&900\\
57180.722189&-19169.61	&1.35&50.47&-5.0233&0.0034&900\\
57181.488048&-19163.85	&1.41&50.17&-5.0279&0.0040&900\\
57181.727141&-19161.05	&1.30&48.62&-5.0192&0.0032&900\\
57182.478739&-19169.18	&1.58&51.14&-5.0251&0.0046&900\\
57182.689347&-19164.93	&1.32&52.94&-5.0279&0.0033&900\\
57183.511190&-19173.15	&1.55&53.09&-5.0158&0.0040&900\\
57183.717099&-19168.87	&1.71&42.67&-5.0186&0.0045&900\\
57184.479947&-19178.99	&3.00&54.98&-5.0108&0.0113&900\\
57184.690358&-19176.01	&2.12&47.09&-5.0221&0.0062&900\\
57185.477951&-19174.57	&1.47&49.48&-5.0271&0.0038&900\\
57185.681916&-19171.39	&1.45&52.06&-5.0264&0.0038&900\\
57186.475909&-19168.82	&1.42&45.79&-5.0353&0.0039&900\\
57186.704226&-19163.21	&1.25&45.28&-5.0338&0.0031&900\\
57188.483535&-19168.47	&1.78&54.30&-5.0352&0.0054&900\\
57188.698634&-19168.32	&1.42&55.20&-5.0352&0.0039&900\\
57189.475993&-19173.50	&2.02&48.24&-5.0248&0.0059&900\\
57189.690663&-19169.02	&1.90&58.09&-5.0305&0.0055&900\\
57190.488312&-19165.22	&1.57&42.69&-5.0422&0.0049&900\\
57190.704741&-19162.65	&1.34&48.74&-5.0361&0.0036&900\\
57191.488650&-19163.93	&1.28&55.45&-5.0351&0.0034&900\\
57191.704385&-19166.76	&1.33&47.40&-5.0252&0.0034&900\\
57192.485794&-19168.26	&1.37&47.73&-5.0348&0.0037&900\\
57192.702373&-19170.10	&1.39&44.88&-5.0258&0.0035&900\\
57193.488920&-19160.15	&1.47&47.02&-5.0268&0.0035&900\\
57193.703220&-19165.44	&1.32&48.50&-5.0358&0.0034&900\\
57195.488609&-19165.45	&1.84&46.56&-5.0371&0.0056&900\\
57195.637281&-19166.29	&1.75&45.87&-5.0322&0.0053&900\\
\enddata
\tablenotetext{a}{The full table is published in the journal's electronic edition. A portion is reproduced here to show its form and content.}
\end{deluxetable}


\end{document}